\documentclass[prd,aps,preprint,tightenlines,showpacs,nofootinbib,superscriptaddress]{revtex4-1}
\usepackage{mathrsfs}
\usepackage{amsfonts}
\usepackage{amsmath}
\usepackage{amssymb}
\usepackage{array}
\usepackage{verbatim}
\usepackage{bm}
\usepackage{epsfig}
\usepackage{graphicx,color}
\usepackage{relsize}
\usepackage{lineno}
\usepackage{float}
\usepackage{multirow}
\RequirePackage{xspace}

\def\pom{{I\!\!P}}

\newcommand{\rb}{\mbox{\boldmath $b$}}
\newcommand{\rr}{\mbox{\boldmath $r$}}
\newcommand{\qb}{\mbox{\boldmath $q$}}

\begin{document}

\title{\boldmath Exclusive processes in $ep$ collisions at the EIC and LHeC: \\ A closer look on the predictions of  saturation models}

\author{Ya-Ping Xie}
\email{xieyaping@impcas.ac.cn}
\affiliation{Institute of Modern Physics, Chinese Academy of Sciences,
	Lanzhou 730000, China}
\affiliation{University of Chinese Academy of Sciences, Beijing 100049, China}

\author{V.~P. Gon\c{c}alves}
\email{barros@ufpel.edu.br}
\affiliation{High and Medium Energy Group, \\
	Instituto de F\'{\i}sica e Matem\'atica, Universidade Federal de Pelotas\\
	Caixa Postal 354, CEP 96010-900, Pelotas, RS, Brazil}
\affiliation{Institute of Modern Physics, Chinese Academy of Sciences,
	Lanzhou 730000, China}

\begin{abstract}

The exclusive production of vector mesons and photons in $ep$ collisions is investigated considering three phenomenological saturation models based  on distinct assumptions for the treatment of the dipole - hadron scattering amplitude. The latest high precision HERA data for the reduced and vector meson cross sections are used to  update the saturation model proposed by Marquet - Peschanski - Soyez (MPS), which predicts that the saturation model is dependent of the squared momentum transfer $t$.  The MPS predictions for the photon virtuality, energy and $t$ - dependencies of the exclusive $\rho$, $J/\Psi$ and DVCS cross sections are presented and a detailed comparison with the results derived  using  the impact parameter saturation models is performed. Our results indicate that a future experimental analysis of the $t$ - distribution $d\sigma/dt$ for exclusive processes in the kinematical range that will covered by the EIC and LHeC, considering the distinct photon polarizations and  large values of $t$, will be able to discriminate between the distinct approaches for the QCD dynamics at high energies.
\end{abstract}

\maketitle

\section{Introduction}

\begin{figure}[t]
	\centering
\includegraphics[width=0.65\textwidth]{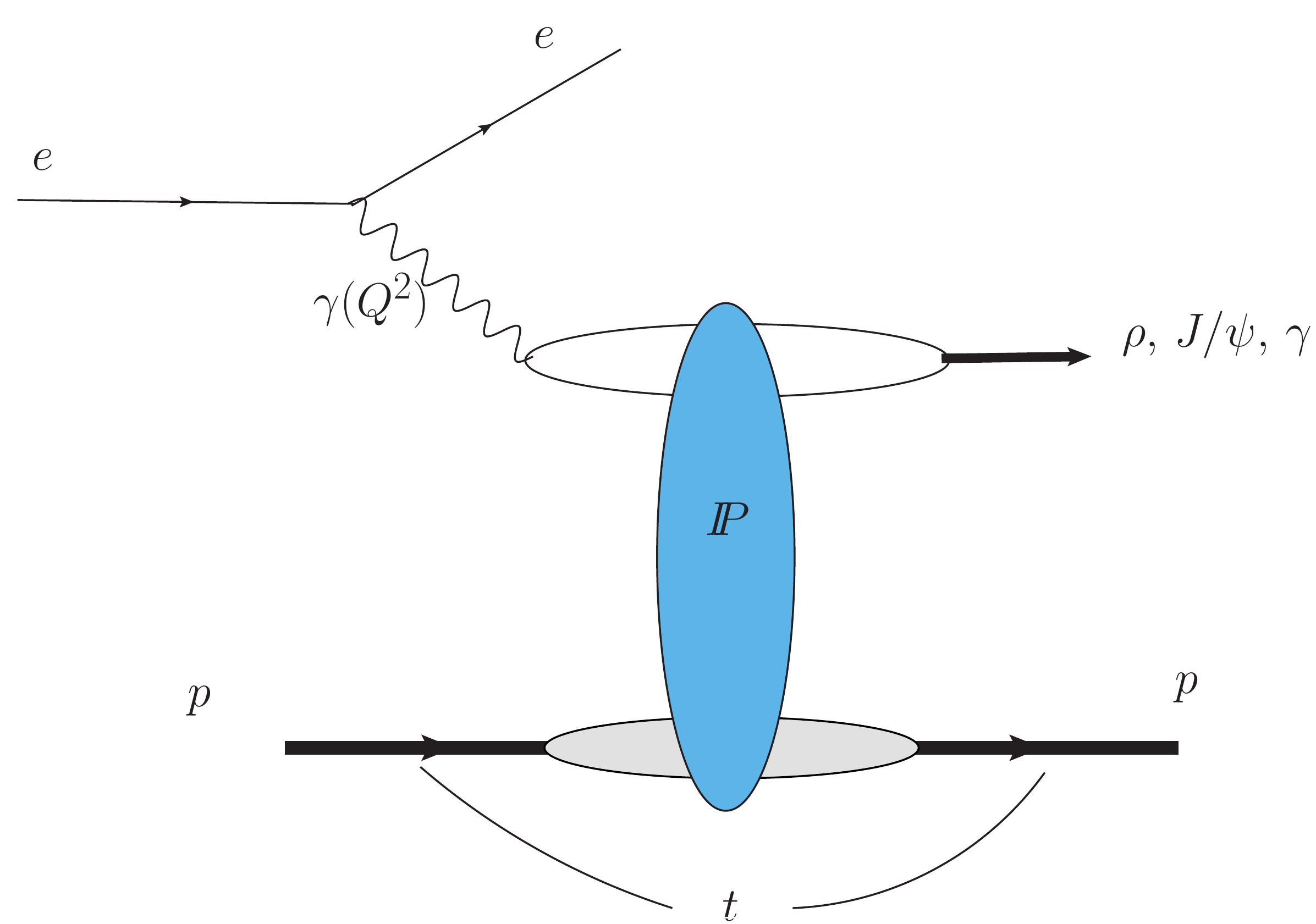}
	\caption{Exclusive processes in $ep$ collisions.}
	\label{fig:dia}		
\end{figure}

The main goal of the future Electron - Ion Colliders at the BNL (EIC) \cite{eic} and at the LHC (LHeC) \cite{lhec} is to achieve a deeper knowledge of the hadronic structure at high energies through the deep 
inelastic scattering (DIS) process, where an electron emits a virtual photon which 
interacts with a proton/nuclear target. In particular, the proton structure can then be studied through 
the $\gamma^{*}p$ interaction, with the behavior of the observables  being determined by the QCD dynamics at high energies. Previous experimental studies,  carried out at HERA,  have shown that the gluon density inside the proton grows with the energy (See e.g. Ref.~\cite{Klein:2008di}). 
Therefore, at high energies, a hadron becomes a dense system and  the non - linear (saturation) effects 
inherent to  the QCD dynamics may become visible \cite{hdqcd}.  Such aspect has motivated the development of an intense phenomenology over the last years, which have demonstrated that diffractive processes, characterized by a rapidity gap in the final state, are the most promising one to study the  high - density regime of QCD and to probe the the gluonic structure of protons and nuclei (See, e.g. Refs. \cite{erike_ea2,vmprc,Caldwell,Lappi_inc,Toll,armestoamir,diego,Lappi:2014foa,Mantysaari:2016ykx,Mantysaari:2016jaz,Diego1,contreras,Luszczak:2017dwf,Mantysaari:2017slo,Diego2,Bendova:2018bbb,cepila,Lomnitz:2018juf,Mantysaari:2019jhh,Hatta:2017cte,Goncalves:2020ywm,Mantysaari:2020lhf,Bendova:2020hkp}). In particular, the exclusive production of a vector meson or a photon, represented in Fig.~\ref{fig:dia},  was demonstrated to be strongly sensitive to the underlying QCD dynamics, since this process is driven by the gluon 
content of the target, with the  cross section being  proportional 
to the square of the dipole - hadron scattering amplitude. Moreover, by measuring the squared momentum transfer $t= - \qb^2$, where $\qb$ is the momentum transfer, the transverse spatial distribution of the gluons in the hadron wave function can also be studied \cite{Mantysaari:2020axf}.

The exclusive processes in $ep$ collisions are characterized by the produced state $E$ ($= \rho,\, J/\Psi, \, \gamma$) and an intact proton in the final state, with a rapidity gap separating these systems. In the color dipole formalism \cite{dip}, the scattering amplitude  can be factorized in terms of the photon wave function $\Psi^{\gamma}$, which describes the  fluctuation of the virtual photon into a $q \bar{q}$ color dipole, the dipole-hadron scattering amplitude $\mathcal{T}$, which describes the scattering by a color singlet exchange ($\pom$),  and  the wave function $\Psi^E$ that describes the $q \bar{q}$ recombination into the exclusive final state $E$. One has that the dipole-hadron scattering amplitude $\mathcal{T}$  can be expressed in terms of the momentum transfer $\qb$ or the impact parameter $\rb$, with both representations being related by a Fourier transform. 

In recent years, several groups \cite{Kowalski:2003hm,Kowalski:2006hc,Watt:2007nr,cepila_nucleon,cepila_nucleon1} have focused on the impact parameter representation and proposed different phenomenological approaches to describe ${\cal{T}}(x,\rr,\rb)$, which are  based on the Color Glass Condensate (CGC) formalism \cite{CGC} and that successfully describe a large set of observables in $ep$, $pp$, $pA$, and $AA$ collisions. In particular, in our analysis we will consider the bSat  and bCGC  approaches \cite{Kowalski:2006hc,Watt:2007nr}, which are able to describe the data for exclusive processes in the kinematical range covered by the HERA experiment. 
A shortcoming of bSat and bCGC models is that the $\rb$ dependencies of the dipole amplitudes are {\it ad hoc} assumptions of the models, which are expected to capture the main contributions associated to non-perturbative physics that  are not  taken into account by the CGC weak-coupling approach. Such aspect has motivated an intense debate in the literature (See, e.g. Refs. \cite{cepila_nucleon1,levinb}). In contrast, the modelling of the scattering amplitude in the momentum transfer representation, ${\cal{T}}(x,\rr,\qb)$, was only discussed in Ref.~\cite{mps} considering the  results derived in Refs. \cite{Marquet:2005qu,Marquet:2005zf} for the exact solutions  of  the Balitsky - Kovchegov (BK) equation \cite{BK} at nonzero momentum transfer. 
Such model, denoted as MPS hereafter,  predicts that the saturation scale becomes proportional to $t$ and that non-perturbative contributions can be factorized when the scattering amplitude is described in the momentum transfer representation, which is not the case in the bSat and bCGC models. One of the goals of this paper is to update the analysis performed in Ref.~\cite{mps} considering the lastest combined HERA data, which  were already used to update the bSat and bCGC models. Another goal is to perform a detailed comparison between the predictions of the MPS, bSat and bCGC models for the 
exclusive $\rho$, $J/\Psi$ and $\gamma$ production  in $ep$ collisions at the kinematical range that will be probed by the EIC and LHeC. In particular, we will present an extensive comparison between the predictions for the exclusive cross sections considering distinct values of the photon virtuality $Q^2$, center - of - mass energy $W$ and squared momentum transfer $t$.  Our main motivation is to verify if the analysis of exclusive processes in future $ep$ colliders could be used to discriminate between these distinct treatments of the QCD dynamics at high energies.

It is important to emphasize that, in recent years, several groups have improved the treatment of exclusive processes 
in the dipole approach, by estimating higher order corrections for the photon impact factor and/or improving the description of the vector meson wave function (See, e.g. Refs. \cite{Lappi:2020ufv,Beuf:2020dxl,Mantysaari:2021ryb}). However, the  proper inclusion of the impact-parameter dependence
of collisions and non - forward corrections in the CGC approach at NLO is  still theme of debate, which motivates the analysis of the existing phenomenological models performed in this paper. 

The paper is organized as follows. In the next section, we present a brief overview of the formalism needed for the description of the exclusive processes in $ep$ collisions and discuss the distinct models for the dipole - proton scattering amplitude employed in our analysis.  In Section~\ref{sec:res}, we exhibit the results for the updated version of the MPS model, obtained by fitting the lastest HERA data, and its predictions are compared with those derived using the bSat and bCGC models. Moreover, we present our predictions for the total cross section and  transverse momentum distribution considering  $ep$ collisions at the EIC and LHeC energies.  Finally, in Section~\ref{sec:conc} we summarize our main conclusions.

\section{Formalism}

In the color dipole formalism, the scattering amplitude for the $\gamma^* p\to E p$ process can be expressed in the momentum  representation  as follows \cite{mps}
\begin{eqnarray}
\mathcal{A}_{T,L}^{\gamma^*p\to Ep}(x, Q^2, t = - \qb^2) = i\int
d^2\rr\int \frac{dz}{4\pi}(\psi_E^*\psi_{\gamma})_{T,L}(z,\rr,Q^2)e^{-iz\qb \cdot \rr}\mathcal{T}(x,\bm r,\qb),
\end{eqnarray}
where $\qb$ is the momentum transfer, $Q^2$ is the photon virtuality and $x = (Q^2 + M^2)/(W^2+Q^2)$, with  $W$ being the center of mass energy of the virtual photon -- proton system and $M$ is the mass of the produced final state. Moreover, 
$r$ is the size of the $q\bar{q}$ dipole  and $z$ and $(1-z)$ are the momentum fractions of the incoming photon momentum carried by the quark and anti-quark, respectively. The overlap  functions  $(\psi_E^*\psi_{\gamma})_{T,L}$  describe the fluctuation 
of a photon with transverse ($T$) or longitudinal ($L$)  polarization into a color dipole and the subsequent formation of the 
final state $E$. In the case of the exclusive photon production, denoted as deeply virtual Compton scattering (DVCS), only the transverse polarization contributes and the
 overlap function $(\psi_{\gamma}^*\psi_{\gamma})$ can be calculated using perturbation theory, being  given by \cite{Kowalski:2006hc}
\begin{eqnarray}
(\psi_{\gamma}^*\psi_{\gamma})_T & = & \sum_f (\psi_\gamma^*\psi_{\gamma})_T^f(z,\rr,Q^2) \nonumber \\
& = &  \sum_f \frac{2N_c}{\pi} \alpha_{em}e_f^2\big\{[z^2+(1-z)^2]\epsilon K_1(\epsilon r)
m_f K_1(m_f r)+m_f^2K_0(\epsilon r)K_0(m_f r)\big\}\,\,,
\end{eqnarray}
where $m_{f}$ and $e_{f}$ are the mass and charge of a quark with flavor $f$ and $\epsilon^2 = z(1-z)Q^2 + m_f^2$.
In contrast, the modelling of the overlap function for the vector meson production  is still a theme of debate, with different models being able to describe e.g. the HERA data. We will assume that  the
vector meson  is  predominantly a quark-antiquark state and that its spin and polarization structures are the same of the  photon \cite{wfbg,wflcg,Forshaw:2003ki,Goncalves:2004bp,Kowalski:2003hm,Kowalski:2006hc}. Such assumptions imply that the transverse and longitudinal overlap functions are given by
\begin{eqnarray}
(\Psi_V^*\Psi_{\gamma})_T & = & \hat{e}_fe\frac{N_c}{\pi z(1-z)}\lbrace  m_f^2
K_0(\epsilon r)\phi_T(r,z)-(z^2+(1-z)^2)\epsilon K_1(\epsilon r)\partial_r
\phi_T(r,z)\rbrace,\notag\\
(\Psi_V^*\Psi_{\gamma})_L & = & \hat{e}_fe\frac{N_c}{\pi}2Qz(1-z)K_0(\epsilon r)\Bigg[M_V\phi_L(r, z)+
\delta \frac{m_f^2-\nabla_r^2}{M_Vz(1-z)}\phi_L(r, z)\Bigg] \,\,,
\end{eqnarray}
where $ \hat{e}_f $ is the effective charge of the quarks,  $N_c = 3$ and $\phi_{T,L}(r,z)$ define the scalar part of the vector meson wave functions.
In this analysis we will consider two popular models employed in the literature -- the Boosted Gaussian (BG)  and Light - Cone Gauss (LCG) -- which differ in the assumptions for the parameter $\delta$ and functions $\phi_{T,L}(r,z)$. In the Boosted Gaussian model, $\delta = 1$ and these functions are expressed as follows
\begin{eqnarray}
\phi_{T,L}(z,r)=\mathcal{N}_{T,L}z(1-z)\exp\Big(-\frac{m_f^2\mathcal{R}^2}{8z(1-z)}
-\frac{2z(1-z)r^2}{\mathcal{R}^2}+\frac{m_f^2\mathcal{R}^2}{2}\Big),
\end{eqnarray} 
where $\mathcal{N}_{T,L}$ and $\mathcal{R}$ are free parameters to be determined by the normalization condition of the wave function and by the decay width. On the other hand,  in the LCG model, $\delta = 0$ and the scalar functions are given by
\begin{eqnarray}
\phi_T(z,r) & = & N_T [z(1-z)]^2 \exp\left(-\frac{r^2}{2R_T^2}\right), \\
\phi_L(z,r) & = & N_Lz(1-z)  \exp\left(-\frac{r^2}{2R_L^2}\right) \,\,.
\end{eqnarray}
As in the BG model, the parameters ${N}_{T,L}$ and $R_{T,L}$ are fixed  by the normalization and  decay width (See e.g. Ref.~\cite{Kowalski:2006hc} for details). The parameters adopted in our analysis are presented in Table \ref{table:WF} for distinct values of the light quark and charm masses. {Some of the
parameters are taken from Refs.~\cite{Kowalski:2006hc,armestoamir}}.
\begin{center}
	\begin{table}[t]
		\begin{tabular}{|c|c|c|c|c|c|c|c|c|c|c|}
			\hline 
			\hline 
			\multicolumn{4}{|c|}{\,}&  \multicolumn{3}{|c|}{{\bf Boosted Gaussian}}& \multicolumn{4}{|c|}{{\bf Light - Cone Gauss}} \tabularnewline
			\hline
		{\bf 	Vector meson}	& $M_V$(GeV) & $m_f$ (GeV) &$\hat{e}_f$& $\mathcal{R}^2(\mathrm{GeV}^{-2})$&  $\mathcal{N}_L$ &$\mathcal{N}_T$ & $R_L(\mathrm{GeV}^{-2})$&$R_T(\mathrm{GeV}^{-2})$ &$N_L$ &$N_T$ \tabularnewline
			\hline
			$\rho$&0.776  &  0.01 & 1/$\sqrt{2}$ &13.3 &0.894 &1.004& 10.4& 21.9& 1.79&5.90 \tabularnewline
			\hline 
			$\rho$ &0.776  &0.14 &1/$\sqrt{2}$ &12.9& 0.853 &0.911&10.4 &21.9 &1.79 &4.47 \tabularnewline
			\hline 
			$J/\psi$  &3.097 &1.27 &2/3& 2.45&0.592&0.596& 3.0&5.6 &0.83 &1.44\tabularnewline
			\hline
			$J/\psi$  &3.097 &1.4 &2/3&2.3 &0.575&0.578&3.0 &6.5 &0.83 &1.23\tabularnewline
			\hline
			\hline 
		\end{tabular}
		\caption{Parameters for the vector meson wave functions used in our analysis ~\cite{Kowalski:2006hc,armestoamir}.}
		\label{table:WF}
	\end{table}
\end{center}

One has that  the energy, photon virtuality and transverse momentum   dependencies of the scattering amplitude for  exclusive processes  are  determined by the evolution of $\mathcal{T}$ and, consequently, are strongly dependent on the description of the QCD dynamics.
At small values of  $x$,  the parton densities increase and the scattering amplitude approaches the unitarity limit, where non - linear effects become important and should be taken into account \cite{hdqcd}. In  order to describe the saturation regime of QCD one can consider the Balitsky - Kovchegov (BK) equation \cite{BK}, which resums QCD fan diagrams in the leading - logarithmic approximation. Such equation has been extensively studied in the literature, mainly considering  that the impact parameter dependence can be factorized, i.e. ${\cal{T}}(x,\rr,\rb) = {\cal{T}}(x,\rr) S(\rb)$ (See, e.g. \cite{Albacete:2010sy}).  It has been shown that the equation for ${\cal{T}}(x,\rr)$ is equivalent to the Fisher - Kolmogorov - Petrovsky - Piscounov (F-KPP) equation \cite{fkpp} and admits asymptotic traveling - wave solutions \cite{Munier:2003vc}, which imply the geometric scale property, ${\cal{T}}(x,\rr) = {\cal{T}}(rQ_s(x))$, that is observed in the HERA data \cite{Stasto:2000er,Goncalves:2003ke,Marquet:2006jb}. In Ref.~\cite{Marquet:2005qu} the authors demonstrated that the traveling - wave method can be extended for the nonforward case when the BK equation is expressed in the momentum  space and that the geometric scaling property can be extended to the case of nonzero momentum transfer. Based on the results obtained in Refs. \cite{Marquet:2005qu,Marquet:2005zf}, the authors have proposed in Ref.~\cite{mps} the phenomenological MPS model for ${\cal{T}}(x,\rr,\qb)$, in which  the scattering amplitude is expressed as follows
\begin{eqnarray}
\mathcal{T}_{MPS}(x,\rr,\qb)=2\pi R_p^2 f(\qb) \times\begin{cases}
\mathcal{N}_0(\frac{rQ_s(x,\qb)}{2})^{2(\gamma_s+(1/\kappa\lambda Y)\ln(2/rQ_s(x,\qb)))},\quad rQ_s(x,\qb)\leqslant 2;\\
1-\exp\big(-a\ln^2(b rQ_s(x,\qb))\big),\quad\qquad rQ_s(x,\qb)>2,
\end{cases}
\end{eqnarray}
where $\mathcal{N}_0$ = 0.7, $Y=\ln(1/x)$ and the constants $a$ and $b$ are given by 
\begin{equation}
\begin{split}
& a=-\frac{\mathcal{N}^2_0\gamma_s^2}{(1-\mathcal{N}_0)^2\ln(1-\mathcal{N}_0)},\\
&b=\frac{1}{2}(1-\mathcal{N}_0)^{-(1-\mathcal{N}_0)/(2\mathcal{N}_0\gamma_s)}.
\end{split}
\end{equation}
One has that the MPS model generalizes the forward saturation model proposed in Ref.~\cite{iim} for the nonforward case and is characterized by a saturation scale that is dependent on the momentum transfer $\qb$ and by the factorization of the form factor $f(\qb)$, which  describes the momentum transfer dependence of the proton vertex. As in Ref.~\cite{mps}, we will assume that $Q_s(x,\qb)=(x/x_0)^{\lambda/2}(1+c\bm q^2)$ and $f(\qb) = e^{-B\bm q^2}$. The parameters $c$, $B$, $R_p$, $\gamma_s$, $\lambda$ and $x_0$ will be obtained by fitting the latest combined HERA data for the reduced and vector meson cross sections (See below).

As discussed in the Introduction, the scattering amplitude can also be estimated in the impact parameter space. In this space, one has that $\mathcal{A}$ is expressed as follows:
\begin{eqnarray}
\mathcal{A}_{T,L}^{\gamma^*p\to Ep}(x, Q^2, t = - \qb^2)= i\int
d^2\rr\int \frac{dz}{4\pi} \int
d^2\rb(\psi_E^*\psi_{\gamma})_{T,L} e^{-i(\rb-(1-z)\rr)\cdot
	\qb} {\cal{T}}(x,\rr,\rb).
\end{eqnarray}
In our analysis we would like to compare the MPS predictions with those derived using the bSat and bCGC models for ${\cal{T}}(x,\rr,\rb)$, which consider distinct assumptions for the impact parameter dependence. One has that  the bSat model assumes  an eikonalized 
form for  the dipole - proton scattering amplitude  that depends on a gluon distribution evolved via DGLAP equation, being given by \cite{Kowalski:2006hc}
\begin{equation}
{\cal{T}}_{bSat}(x,\rr,\rb) = 2\left\{1-\exp \left[-\frac{\pi^2}{2N_c}\rr^2\alpha_s(\mu^2)xg(x, \mu^2)T_p(\rb)\right]\right\},
\end{equation}
where $T_p(\rb)= ({1}/{2\pi B_p})  \exp(-\rb^2/2B_p)$ is the profile function. On the other hand, in the bCGC model, the saturation scale is assumed to be dependent of the impact parameter and the  dipole - proton scattering amplitude is  written as \cite{Watt:2007nr}
\begin{eqnarray}
{\cal{T}}_{bCGC}(x,\rr,\rb)=2\times\begin{cases}
\mathcal{N}_0(\frac{rQ_s}{2})^{2(\gamma_s+(1/\kappa\lambda Y)\ln(2/rQ_s))},\quad\! rQ_s\le2,\\
1-\exp\big(-a\ln^2(b rQ_s)\big),\quad\quad rQ_s>2,
\end{cases}
\end{eqnarray}
with $Q_s(x,\rb)=(x/x_0)^{\lambda/2}\exp(-\frac{\rb^2}{4\gamma_sB_p})$. In our analysis using the bSat and bCGC models, we will consider the parameters determined  in Refs.~\cite{Rezaeian:2012ji, Rezaeian:2013tka} using the latest HERA data and assuming $m_f = 0.01$ GeV for light quarks and $m_c = 1.27$ GeV for the charm. 

Finally, the differential cross section for the exclusive processes will be calculated including the corrections associated to the real part of the amplitude and the skewedness factor, which is related to the fact that the gluons emitted from the quark and antiquark into the dipole can carry different
momentum fractions \cite{Shuvaev:1999ce}. As a consequence, one has that
\begin{eqnarray}
\frac{d\sigma^{\gamma^* p\to E p}}{dt}
=\frac{R_g^2(1+\beta^2)}{16\pi}|\mathcal{A}_{T,L}(x,Q^2,\qb)|^2,
\label{dsdtx}
\end{eqnarray}
where   $\beta$ is the ratio of real and imaginary parts of the amplitude, which will be estimated using that \cite{Kowalski:2006hc}
\begin{equation}
\beta=\tan(\frac{\pi}{2}\bar{\delta}),  \hspace{2cm}      \mbox{with}  \hspace{2cm}
\bar{\delta}=\frac{\partial \ln (\mathrm{Im}\mathcal{A}_{T,L})}{\partial
	\ln1/x}.
\end{equation}
Moreover, the skewedness factor $R_g^2$ will be calculated as follows \cite{Kowalski:2006hc}
\begin{equation}
R_g=\frac{2^{2\bar{\delta}+3}}{\sqrt{\pi}}\frac{\Gamma(\bar{\delta}+5/2)}{\Gamma(\bar{\delta}+4)},
\end{equation}
where $\bar{\delta}$ is defined as above. 
{Adopting a given dipole amplitude and overlap functions, one can obtain the differential cross sections as a function of $W$, $Q^2$ and $t$. After integrating over $t$ the differential cross sections, one can obtain the total cross sections as a function of $W$ and $Q^2$.  }

%\textit{In previous analysis \cite{mps}, the skewdness effect was not considered for exclusive vector meson cross sections. However, in other papers\cite{Kowalski:2006hc,Rezaeian:2012ji,Rezaeian:2013tka}, this effect is taken into account for the exclusive  cross sections. Consequently, we consider this effect for differential cross sections of vector mesons and photons in this work.}

\section{Results}
\label{sec:res}
First of all, we will update the MPS model considering the latest high precision combined HERA data for the reduced cross sections \cite{H1:2015ubc} as well
 as the H1 and ZEUS data for the exclusive $\rho$ and $J/\Psi$ production\cite{ZEUS:1998xpo,H1:1999pji,ZEUS:2007iet,H1:2009cml,H1:2005dtp,ZEUS:2004yeh}. 
 Such reanalysis of available data is motivated by the fact that a large amount of data, with extremely small error bars, were released after the study performed
 in Ref.~\cite{mps}. We will include in our analysis the experimental data for the reduced cross section in the range $Q^2 \in [0.045, 45]$ GeV$^2$ and $x \le 0.01$  and the statistical and systematic experimental uncertainties will be added in quadrature in the calculation of $\chi^2$. 
{Moreover, we also will consider the experimental data for  exclusive vector meson cross sections characterized by  $Q^2 \ge 1$ GeV$^2$. }
 We will consider the Boosted Gaussian (BG) and Light - Cone Gauss (LCG) models for the vector meson 
wave functions and will use the data to fit the free parameters  $c$, $B$, $R_p$, $\gamma_s$, $\lambda$ and $x_0$.  The analysis of the MPS model will 
be performed assuming $m_f = 0.14$ GeV for light quarks and $m_c = 1.4$ GeV for the charm quark.
 
Our results are presented in Table \ref{table:HERA}. One has that the values of $R_p$, $\gamma_s$, $\lambda$ and $x_0$ are similar for the different models of wave functions, which is associated to the fact that they are mainly constrained by the data for the reduced cross section. In contrast, the  $\chi^2$ results of the fit are strongly dependent of the models assumed to describe the vector meson wave functions. One has that the smaller $\chi^2$ is obtained for the LCG model, with this value being smaller than those obtained in Refs. \cite{mps,Rezaeian:2012ji, Rezaeian:2013tka}. In what follows, we will restrict our analysis to the LCG model. It is important to emphasize that one also has considered the possibility that the mesons are described by distinct wave functions, as in Ref.~\cite{mps}, but the resulting $\chi^2$ is not better that the value obtained assuming the LCG model for both mesons.

\begin{center}
	\begin{table}[t]
		\begin{tabular}{|c|c|c|c|c|c|c|c|}
			\hline 
			\hline 
			&$R_p$ (GeV$^{-1}$)& $\gamma_s$ &$\lambda$ &$x_0$& $c$ (GeV$^{-2}$)   & $B$ (GeV$^{-2}$) & $\chi^2/d.o.f$ \tabularnewline
			\hline
			BG & 3.44 $\pm$ 0.06 & 0.720 $\pm$ 0.001&0.207$\pm$ 0.028  &7.97$\times$10$^{-6}\pm$ 0.03$\times$10$^{-6}$ & 1.35 $\pm$ 0.10 & 3.25  $\pm$ 0.02 &  1.39 \tabularnewline
			LCG & 3.46 $\pm$ 0.03 & 0.728 $\pm$ 0.001 & 0.205$\pm$ 0.008  &8.01$\times$10$^{-6}\pm$ 0.03$\times$10$^{-6}$ & 1.48 $\pm$ 0.04 & 3.09  $\pm$ 0.54 &  1.13 \tabularnewline
%		BL & 3.44 $\pm$ 0.07 & 0.721 $\pm$ 0.014&0.207$\pm$ 0.004  &8.04$\times$10$^{-6}\pm$ 0.15$\times$10$^{-6}$ & 1.31 $\pm$ 0.44 & 3.19  $\pm$ 0.44 &  1.13 \tabularnewline
			\hline 
		\end{tabular}
		\caption{Parameters of the MPS model determined from fits to the reduced cross section from the combined HERA data in neutral current unpolarized $ep$ scattering and to the data for the exclusive $\rho$ and $J/\Psi$ production from H1 and ZEUS Collaborations. Results derived using the Boosted Gaussian (BG) and Light - Cone Gauss (LCG) models for the vector meson wave functions. }
		\label{table:HERA}
	\end{table}
\end{center}

\begin{figure}[t]
	\centering
\includegraphics[width=0.48\textwidth]{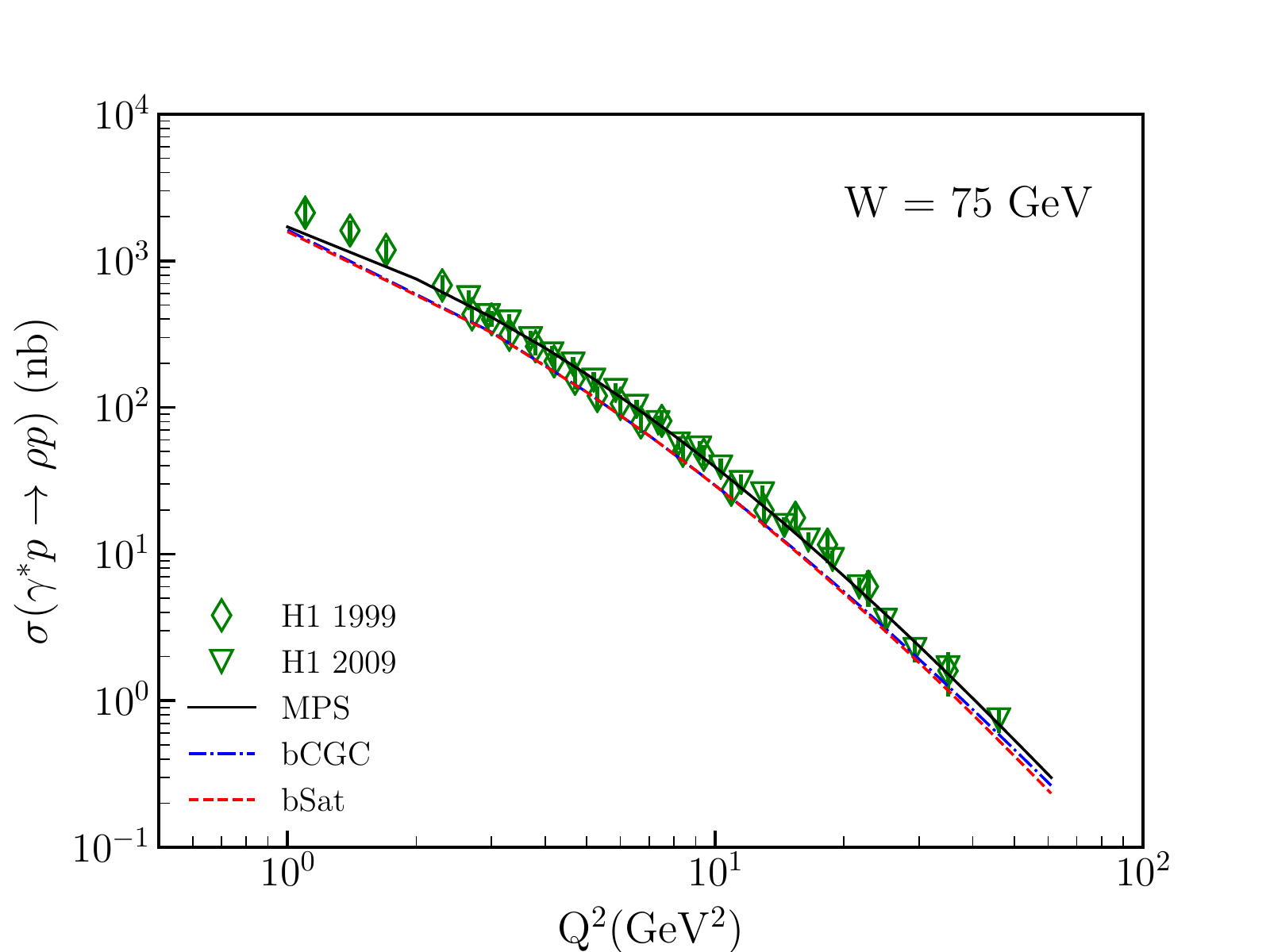}
\includegraphics[width=0.48\textwidth]{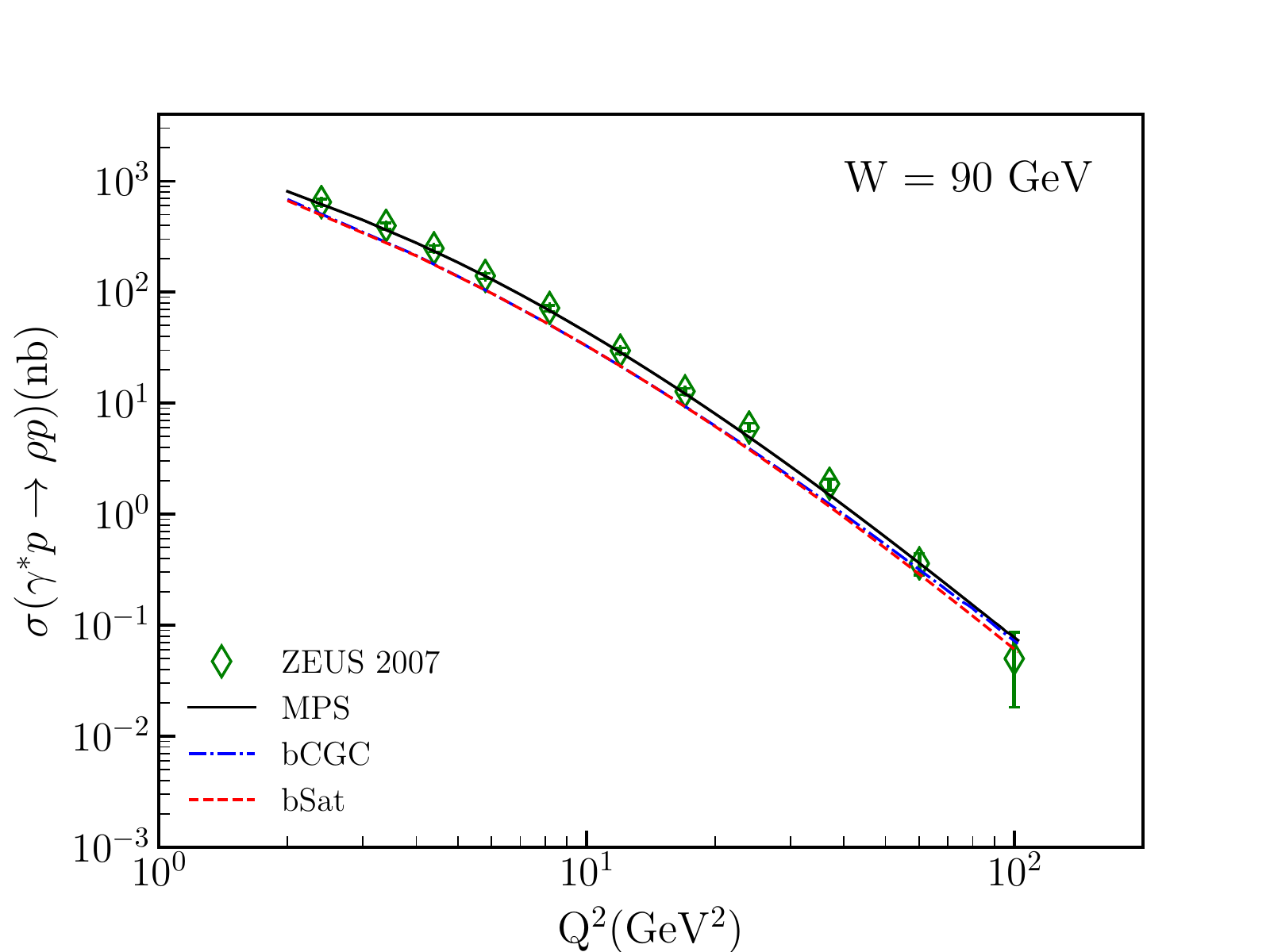} \\
	\includegraphics[width=0.48\textwidth]{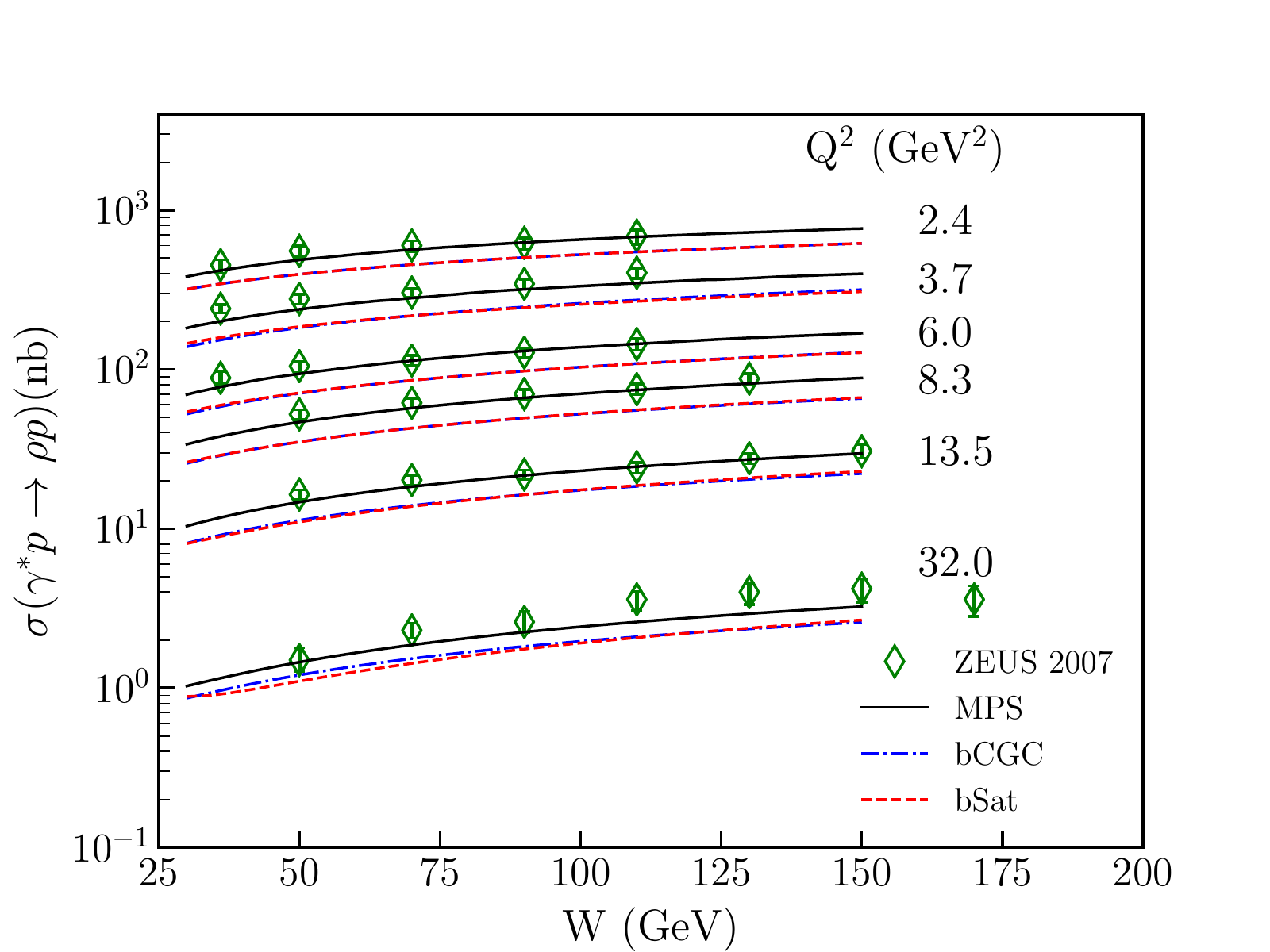}
	\includegraphics[width=0.48\textwidth]{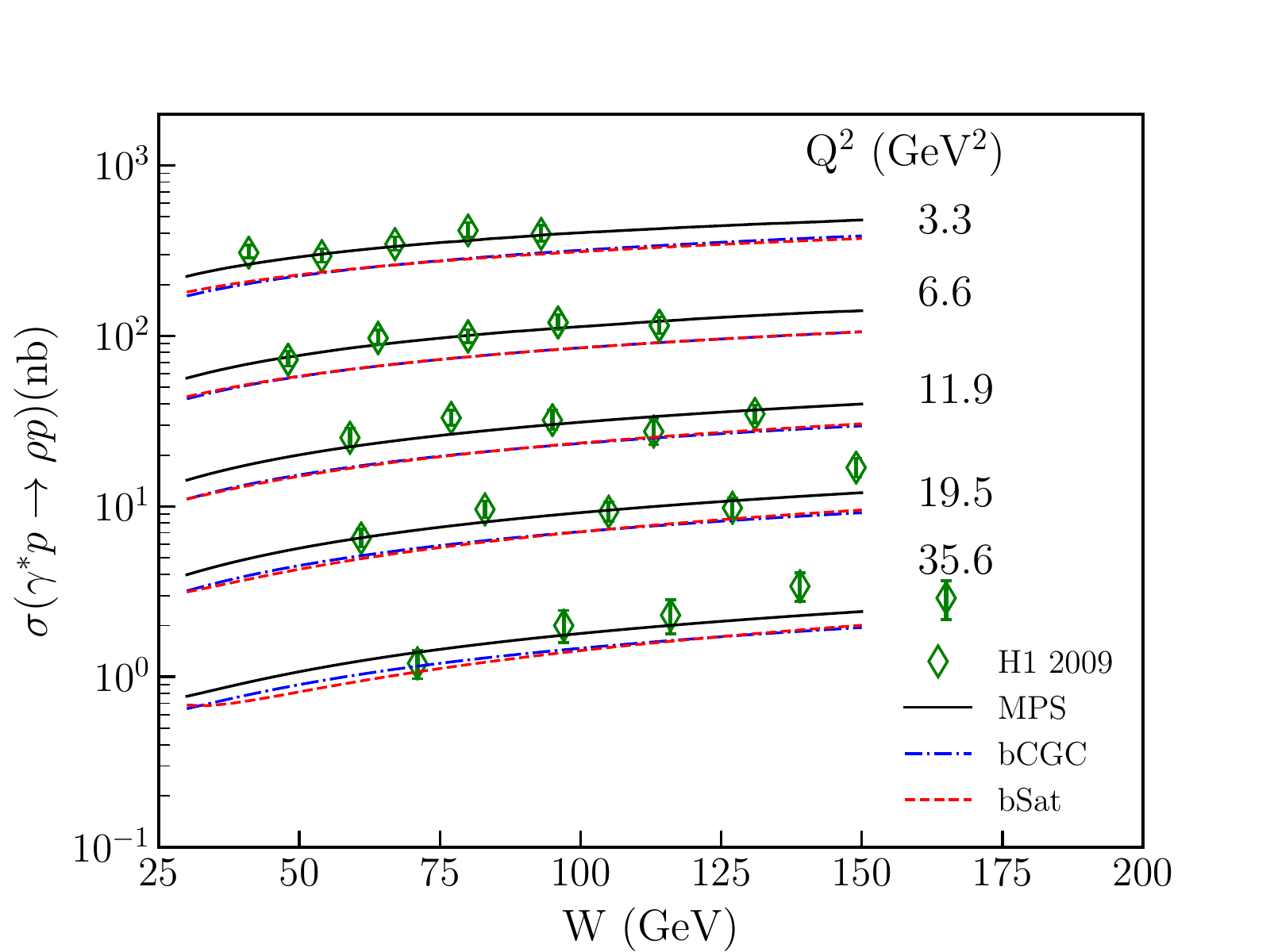} \\
	\includegraphics[width=0.48\textwidth]{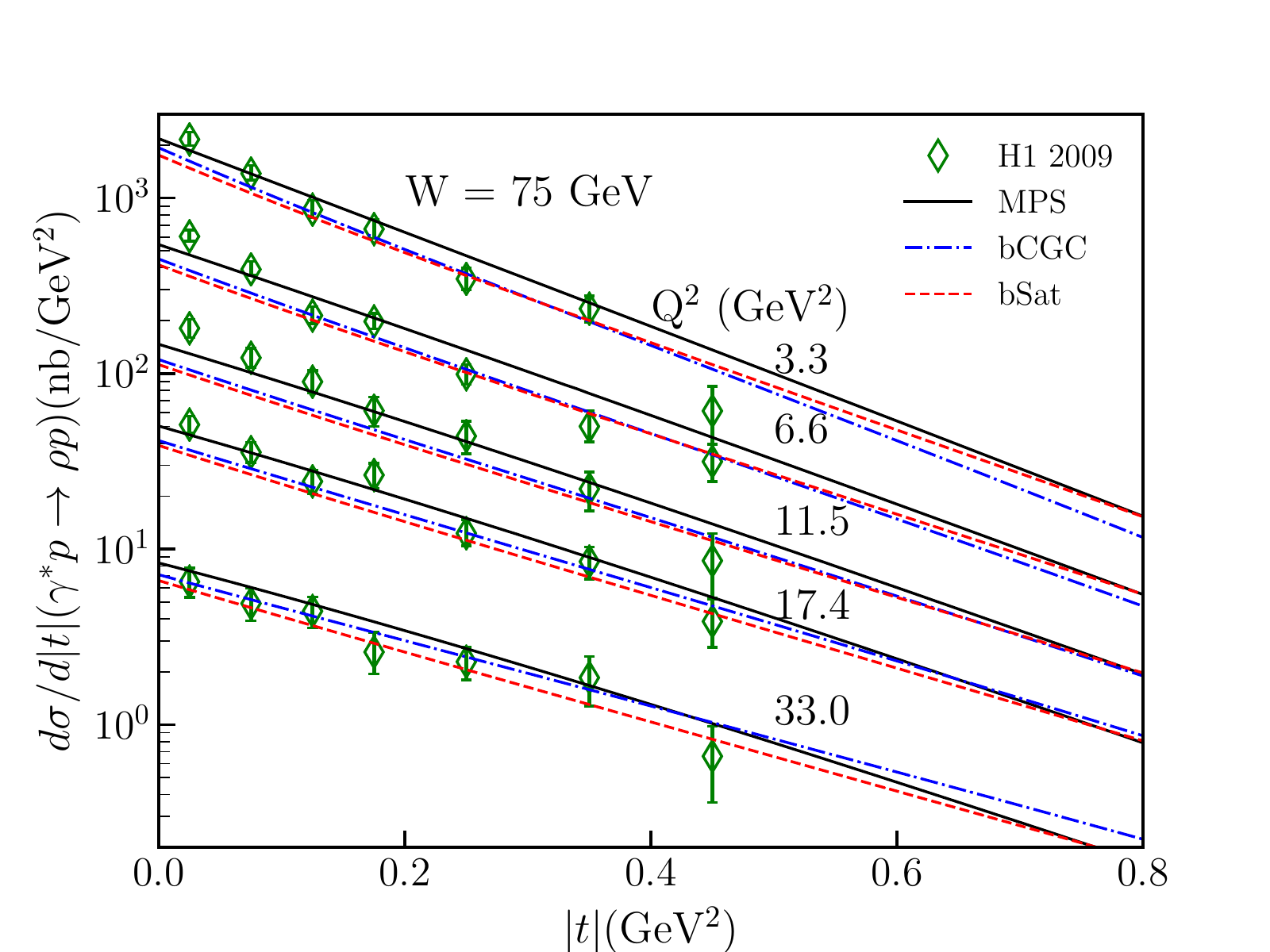}
\includegraphics[width=0.48\textwidth]{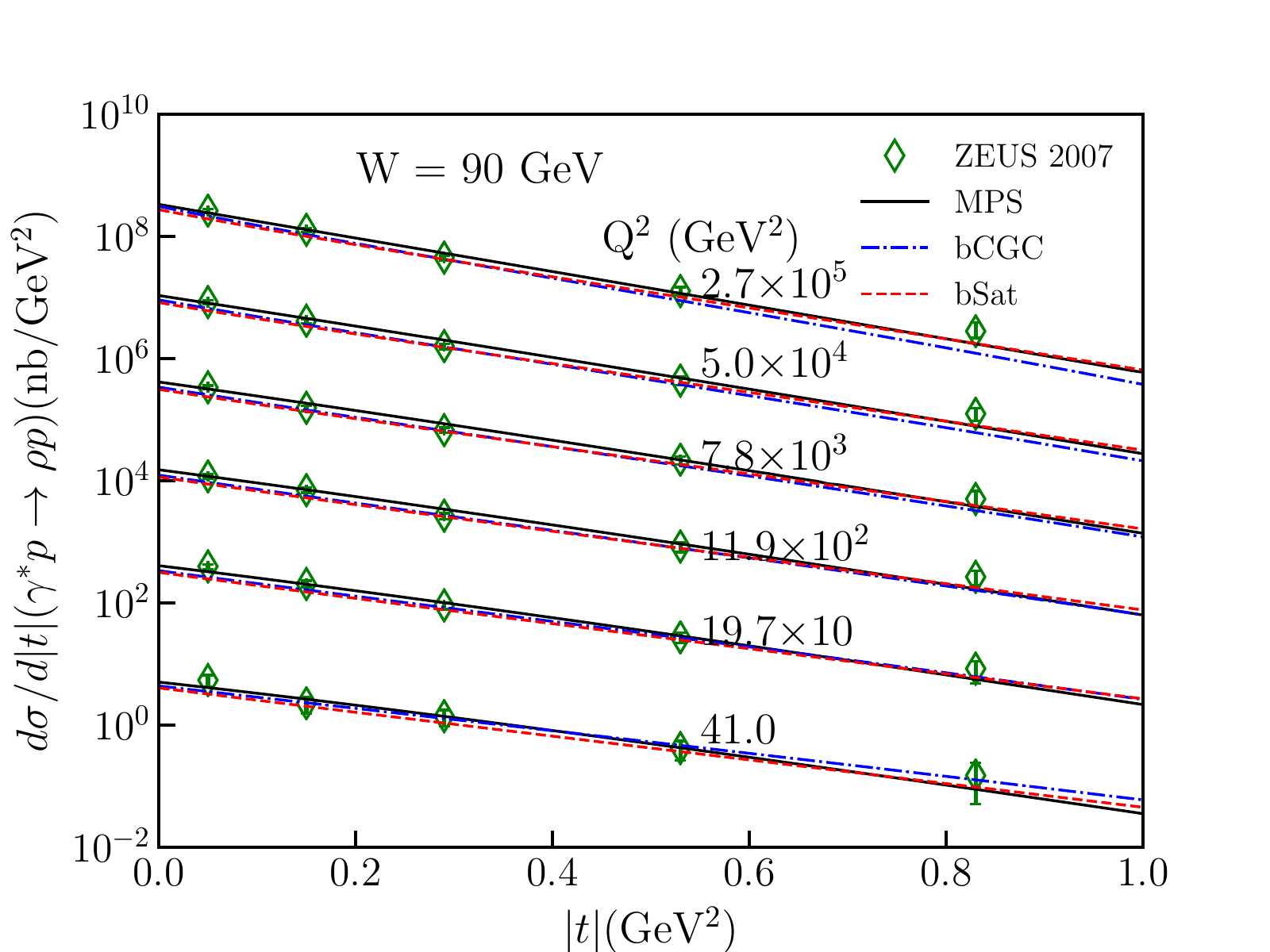}

	\caption{Comparison of the MPS predictions with  the experimental HERA data for the exclusive $\rho$ production. For completeness, the bSat and bCGC results  are also presented. Data from Refs. \cite{H1:1999pji,ZEUS:2007iet,H1:2009cml}.}
	\label{fig:rho_hera}		
\end{figure}

\begin{figure}[t]
	\centering
	\includegraphics[width=0.48\textwidth]{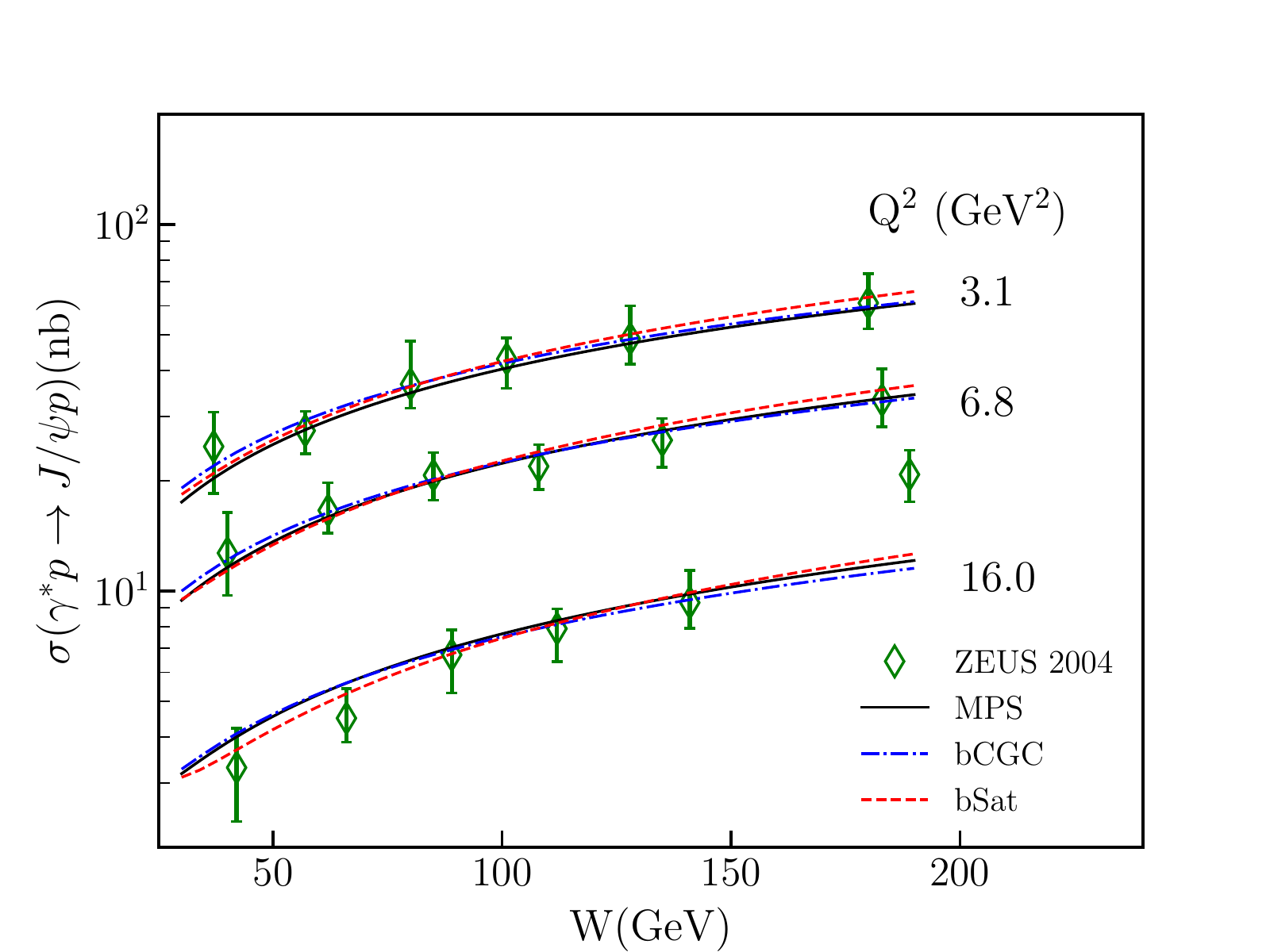}
	\includegraphics[width=0.48\textwidth]{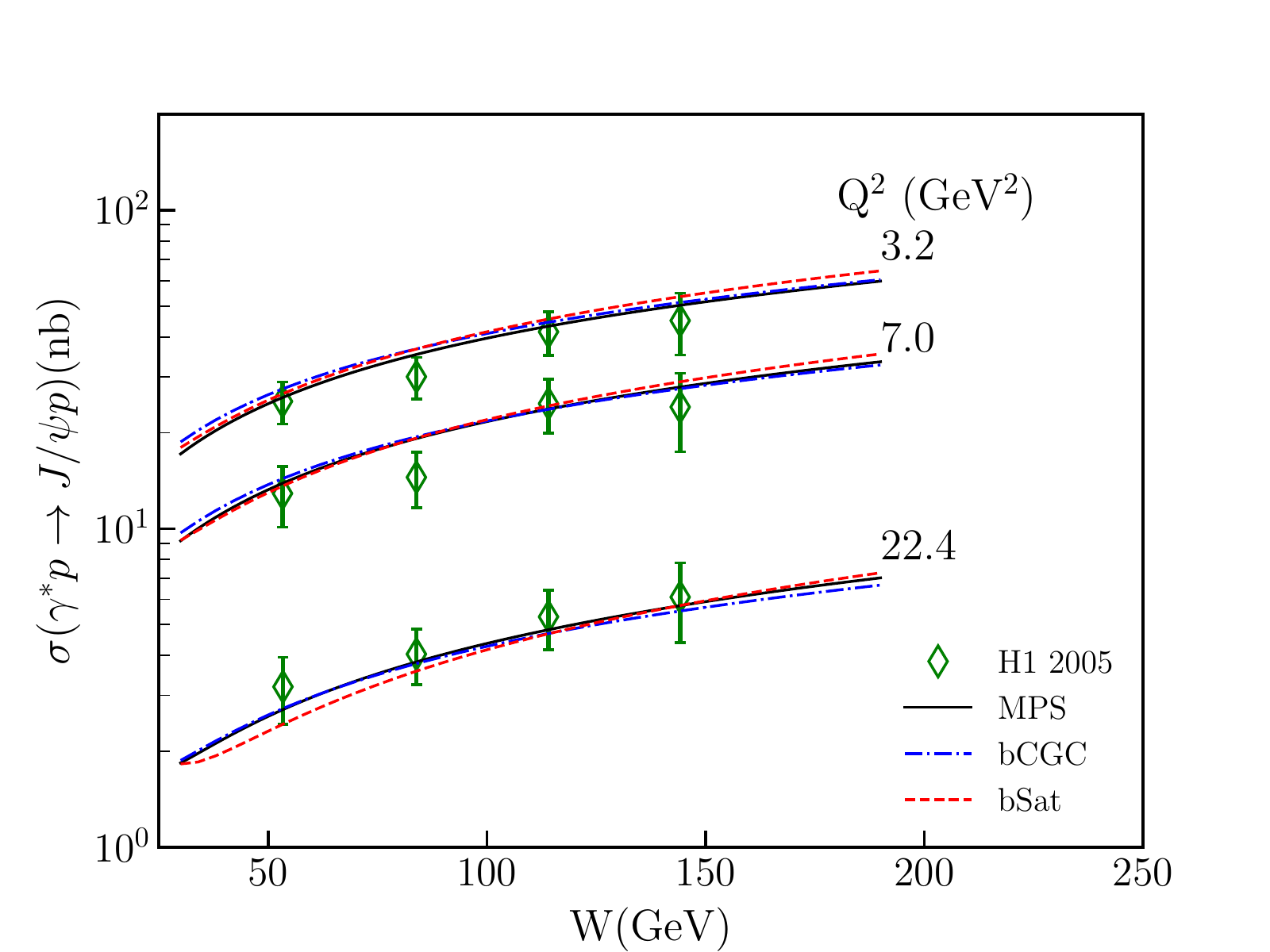} \\
		\includegraphics[width=0.48\textwidth]{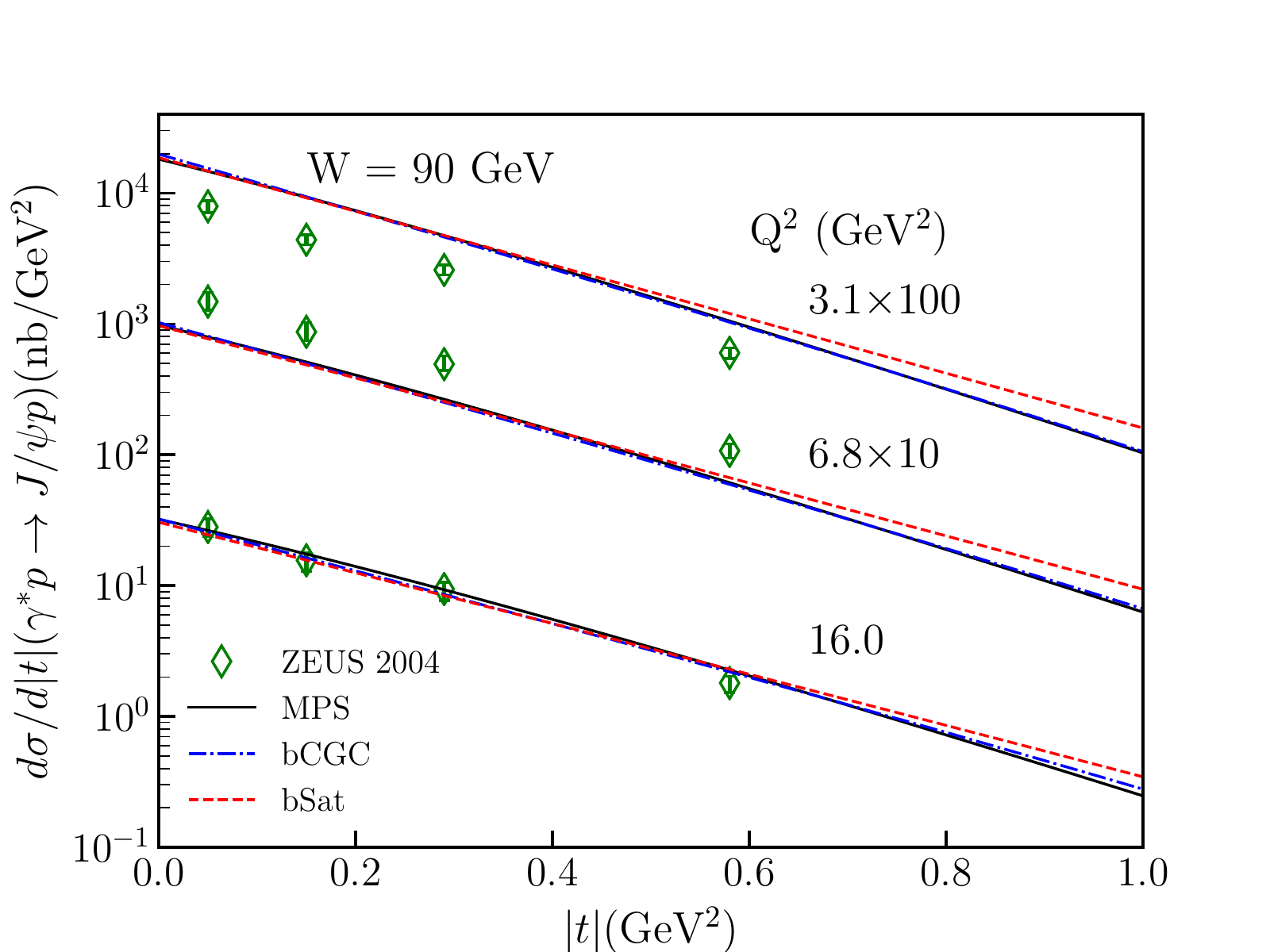}
	\includegraphics[width=0.48\textwidth]{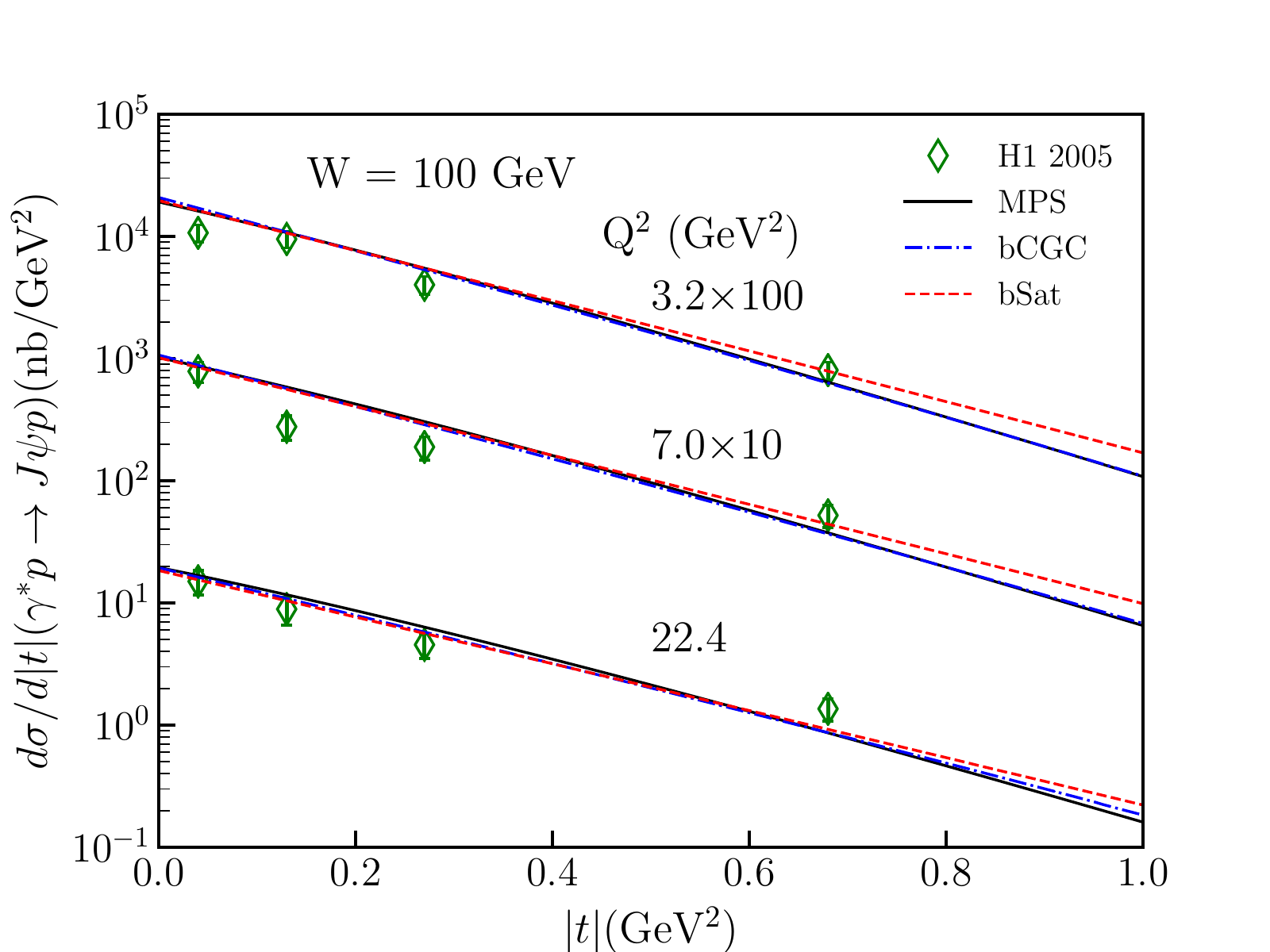}

	\caption{Comparison of the MPS predictions with  the experimental HERA data for the exclusive $J/\Psi$ production. For completeness, the bSat and bCGC results are also presented. Data from Refs. \cite{ZEUS:2004yeh,H1:2005dtp}.}
	\label{fig:jpsi_hera}		
\end{figure}

In Figs. \ref{fig:rho_hera} and \ref{fig:jpsi_hera} we compare the MPS predictions with the experimental HERA data for the exclusive $\rho$ and $J/\Psi$ production used in the fitting procedure. 
For completeness, the bSat and bCGC results are also presented.
One has that in the kinematical range covered by HERA, the predictions of the distinct saturation models for the $J/\Psi$ production are similar, with the main difference occuring in the behavior of $d\sigma/dt$ for  $t \geq 0.4$ GeV$^2$. In contrast, for the $\rho$ production, the MPS model predicts larger values for the total cross section and for the differential distribution at small - $t$.

\begin{figure}[t]
	\centering
	\includegraphics[width=0.48\textwidth]{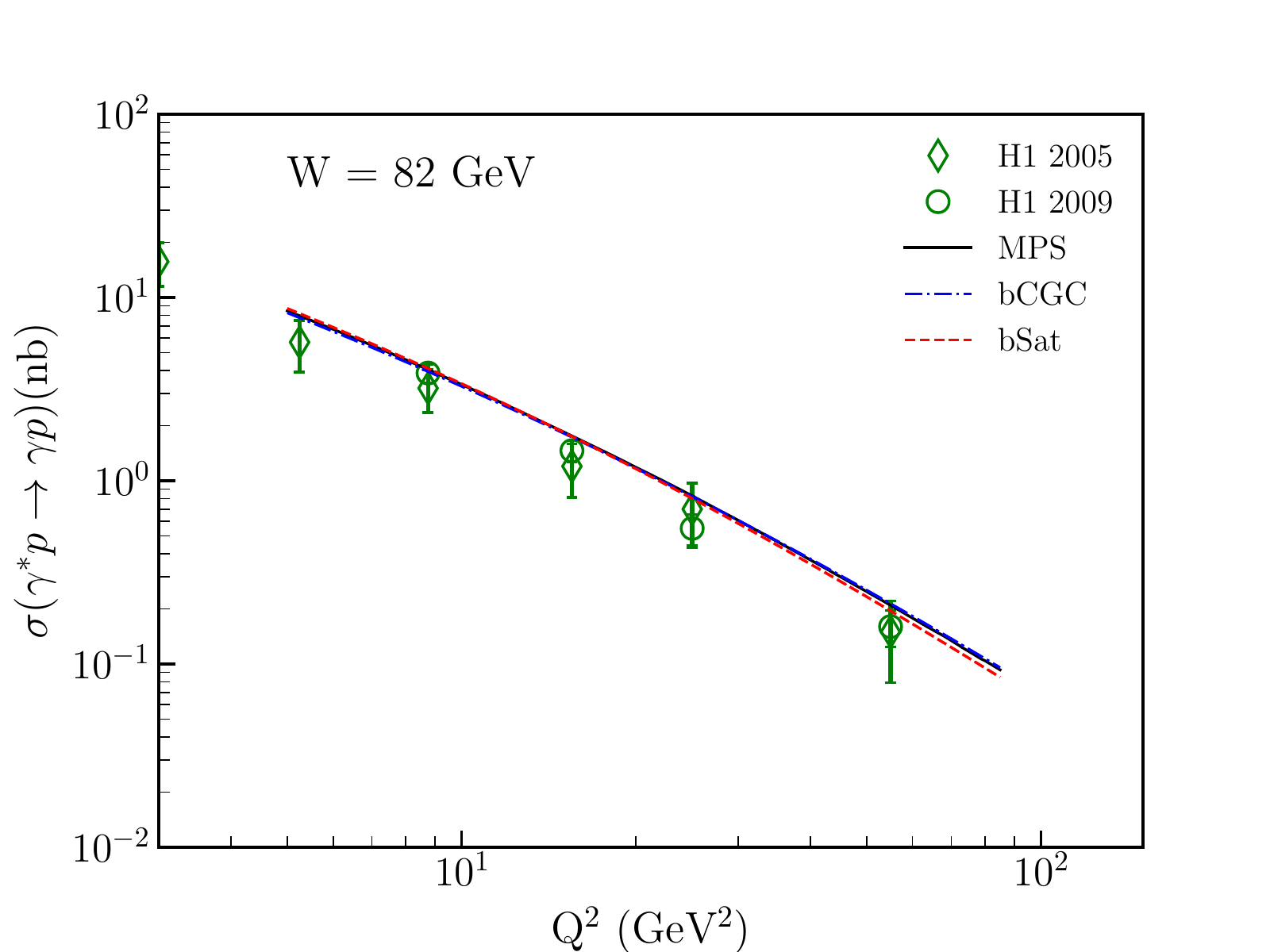}
	\includegraphics[width=0.48\textwidth]{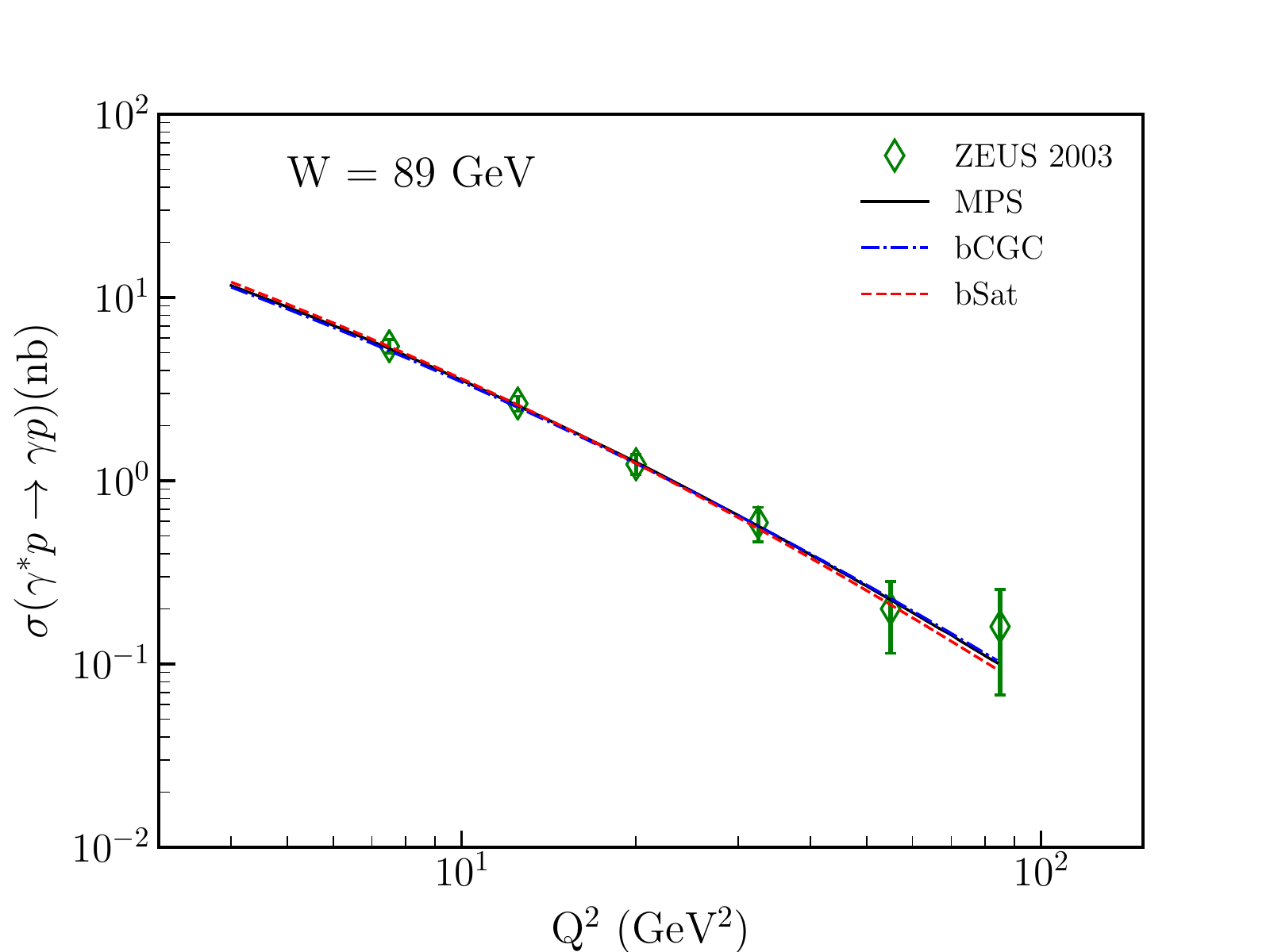} \\
		\includegraphics[width=0.48\textwidth]{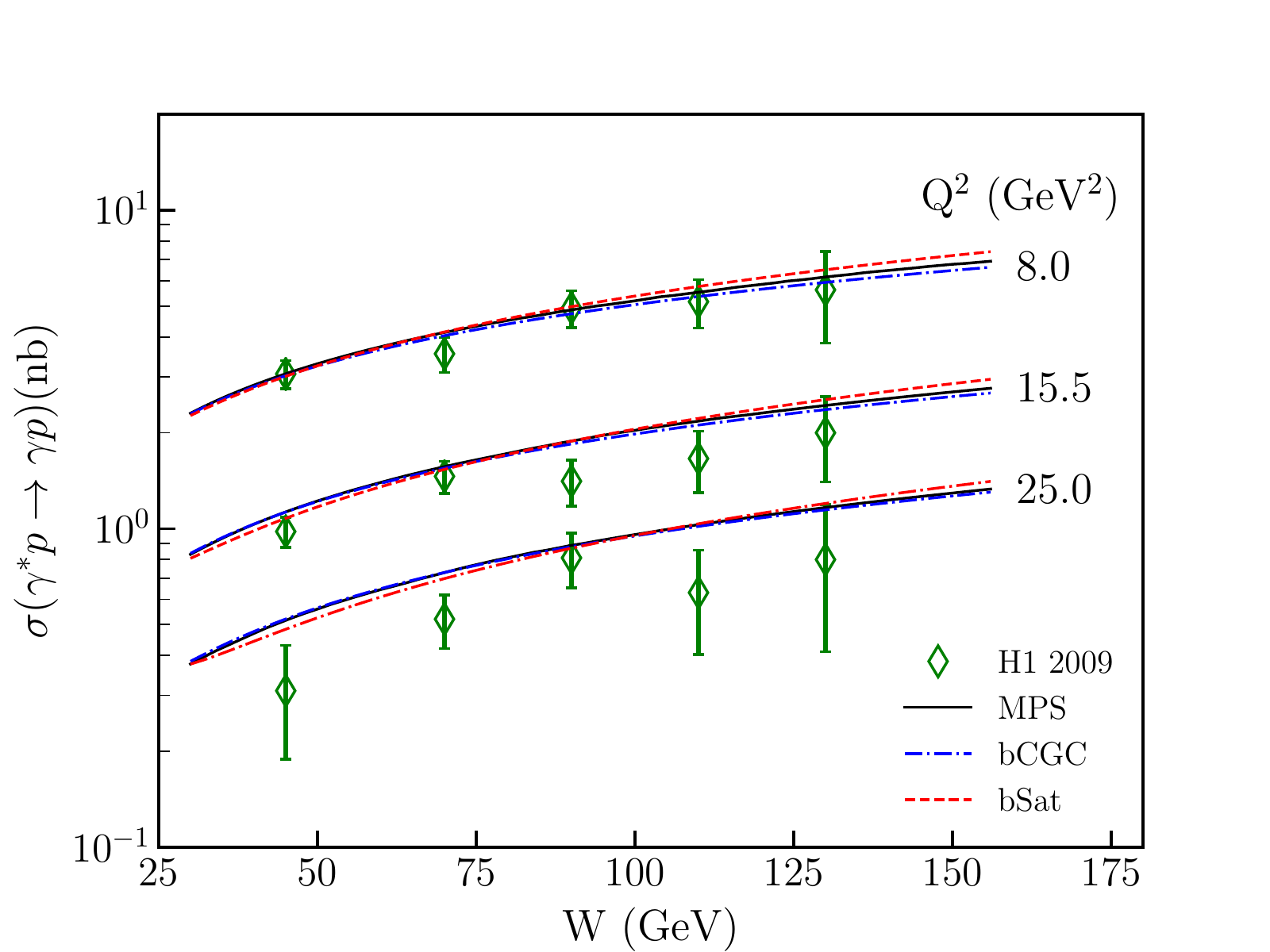}
	\includegraphics[width=0.48\textwidth]{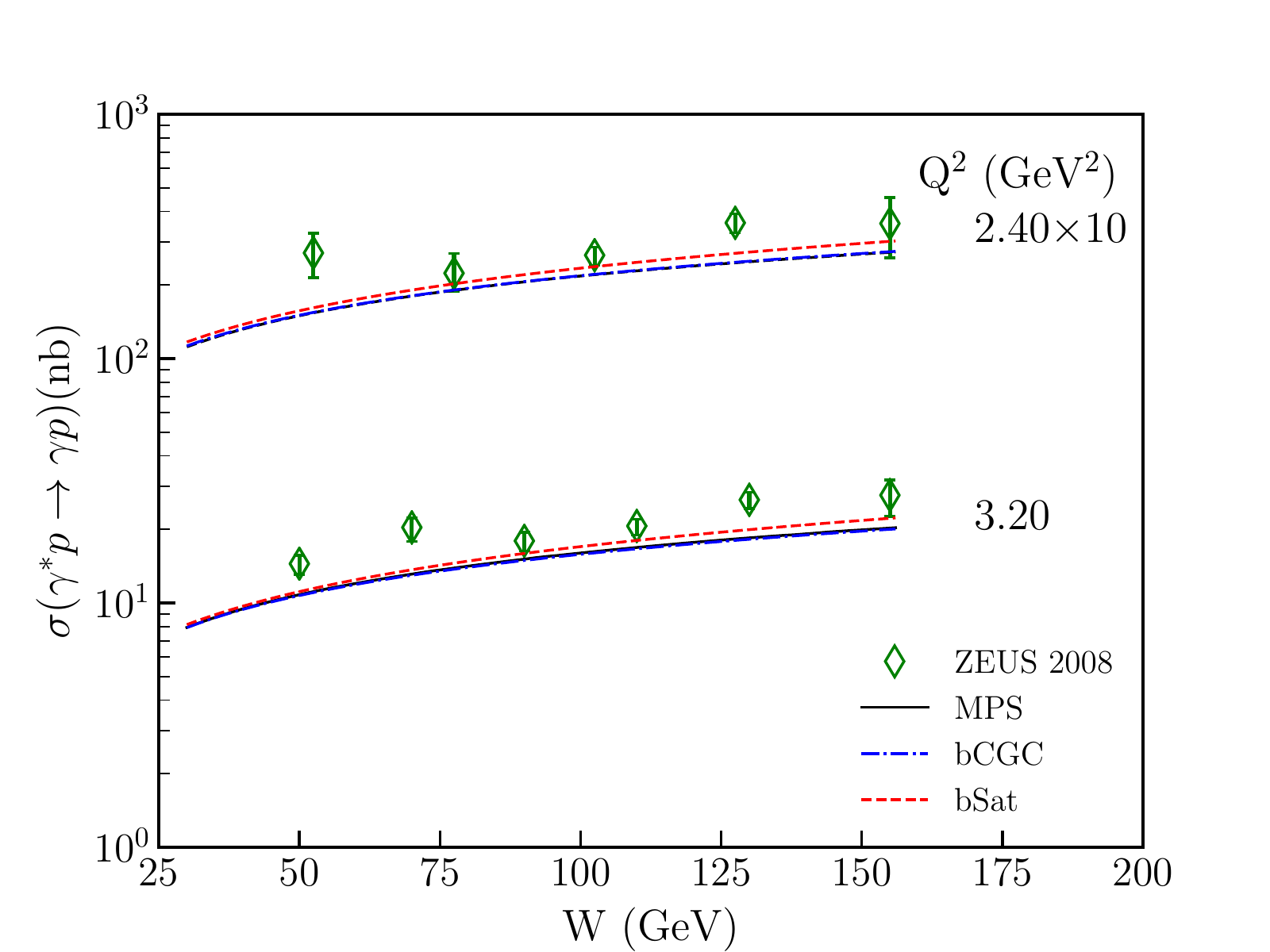} \\
			\includegraphics[width=0.48\textwidth]{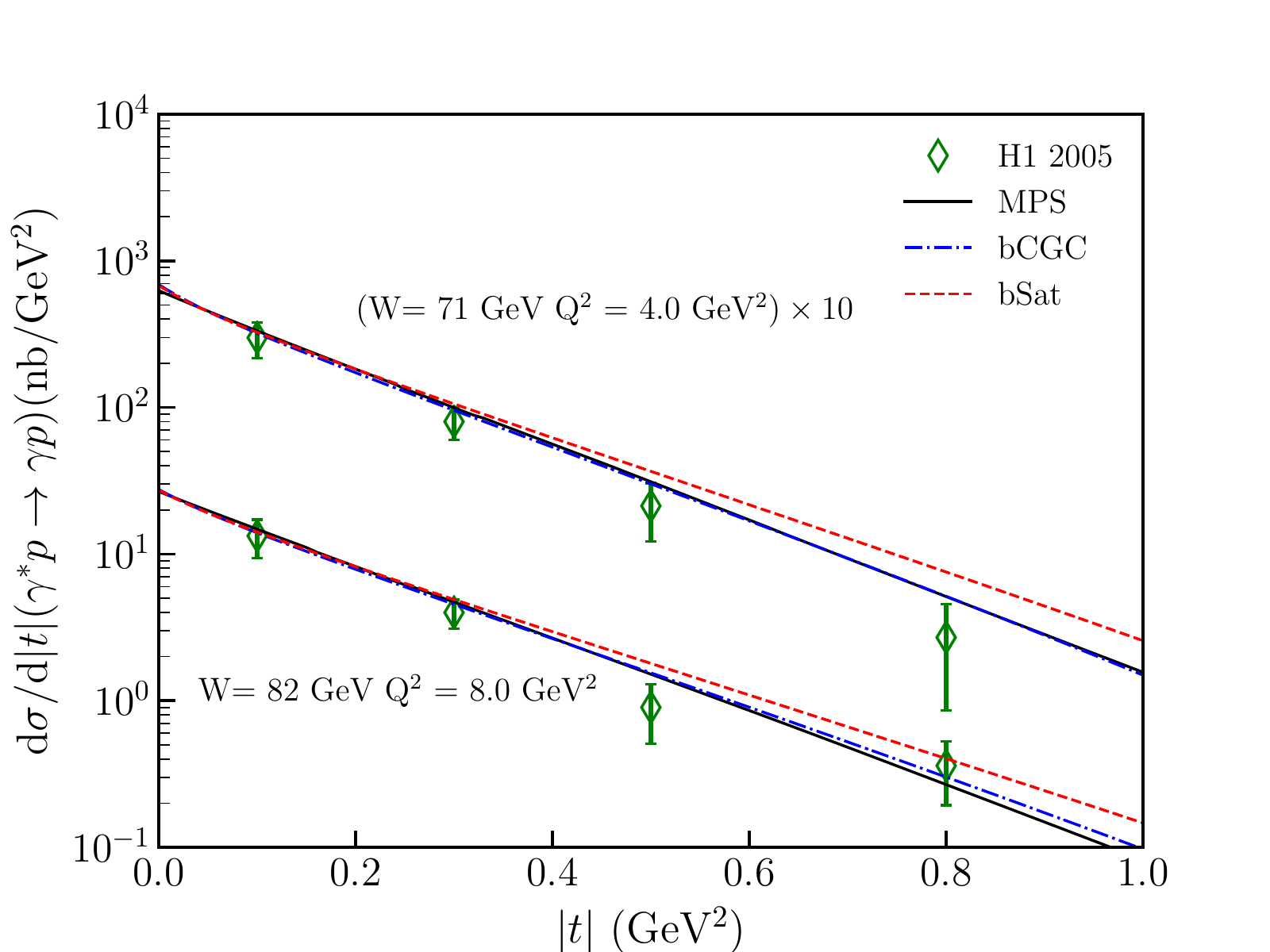}
		\includegraphics[width=0.48\textwidth]{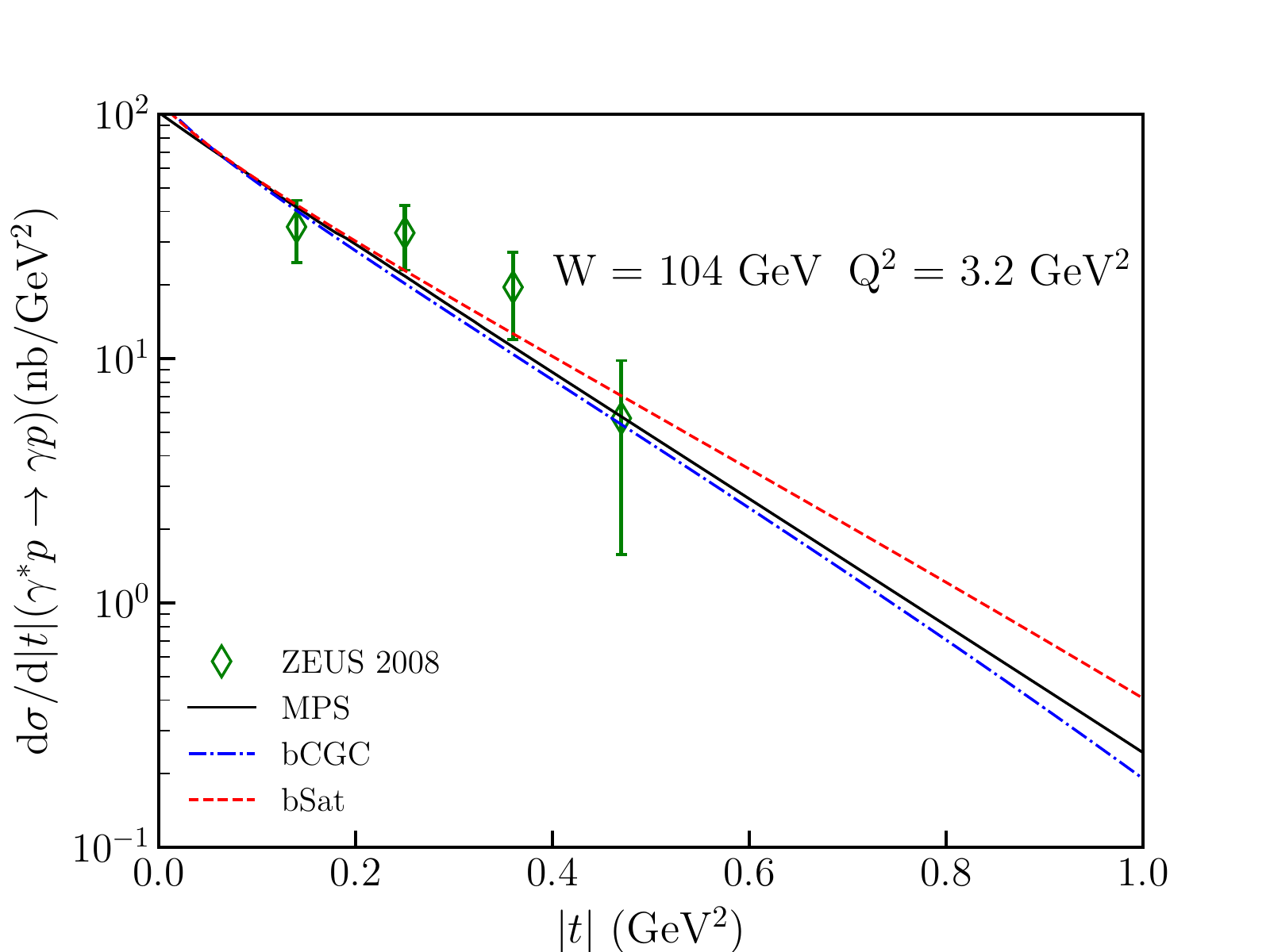}
	\caption{Predictions of the MPS model for the DVCS process considering the kinematical range covered by HERA. For completeness, the bSat and bCGC results are also presented. Data from H1 and ZEUS Collaborations \cite{H1:2005gdw,H1:2009wnw,ZEUS:2003pwh,ZEUS:2008hcd}.}
	\label{fig:dvcs_hera}		
\end{figure}

In order to further test the MPS model, in Fig.~\ref{fig:dvcs_hera} we confront experimental data from H1 and ZEUS Collaborations for the DVCS process with model  predictions for 
photon virtuality, energy and $t$ - dependencies. It is important to emphasize that these data were not included in the fit. One has that the description of the data is quite good, with the MPS predictions being similar to those from the bSat and bCGC models. As in the case of the vector meson production, the predictions of the distinct saturation models differ for $t \geq 0.4$ GeV$^2$.

\begin{figure}[t]
	\centering
	\includegraphics[width=0.32\textwidth]{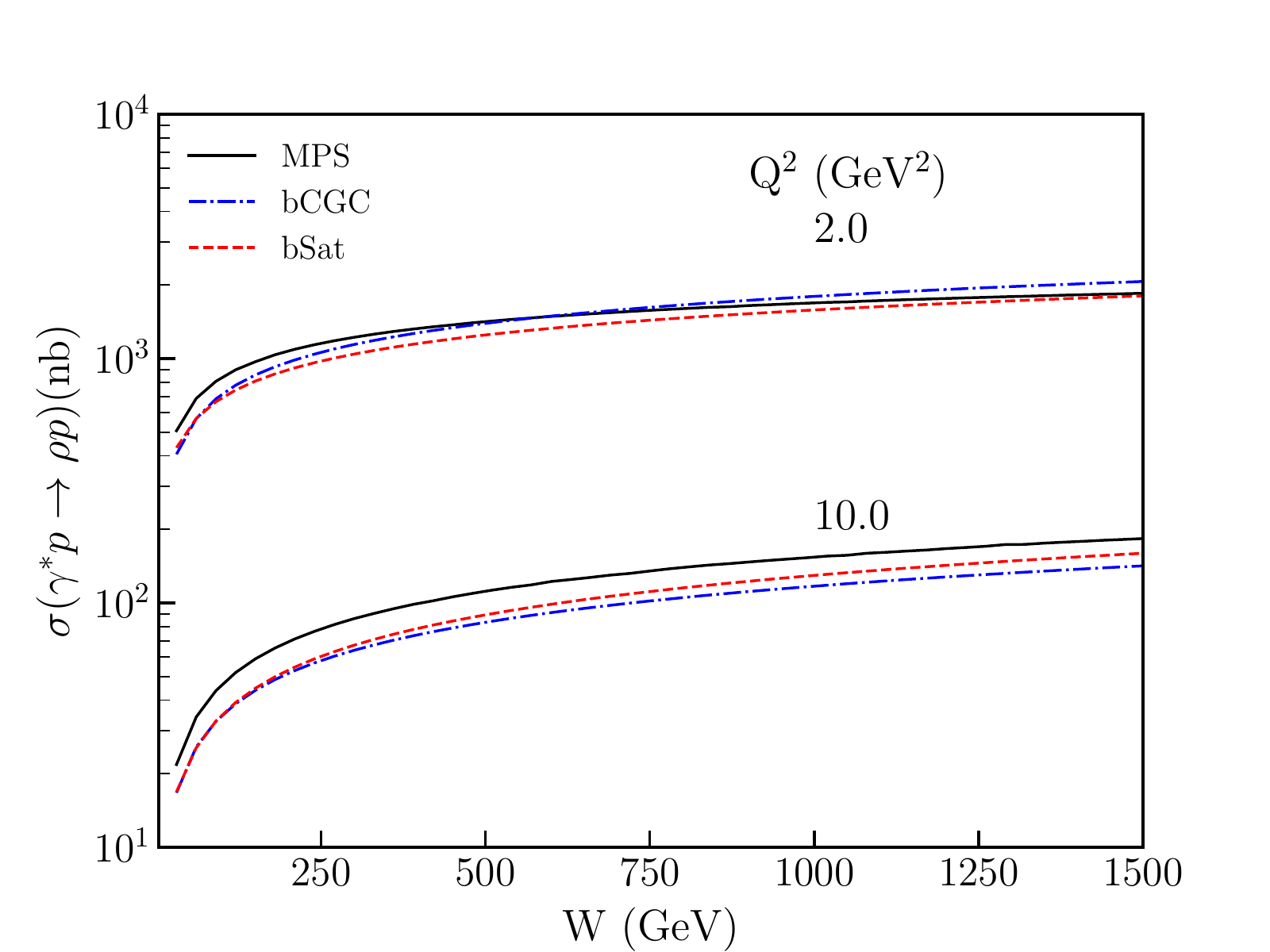}
	\includegraphics[width=0.32\textwidth]{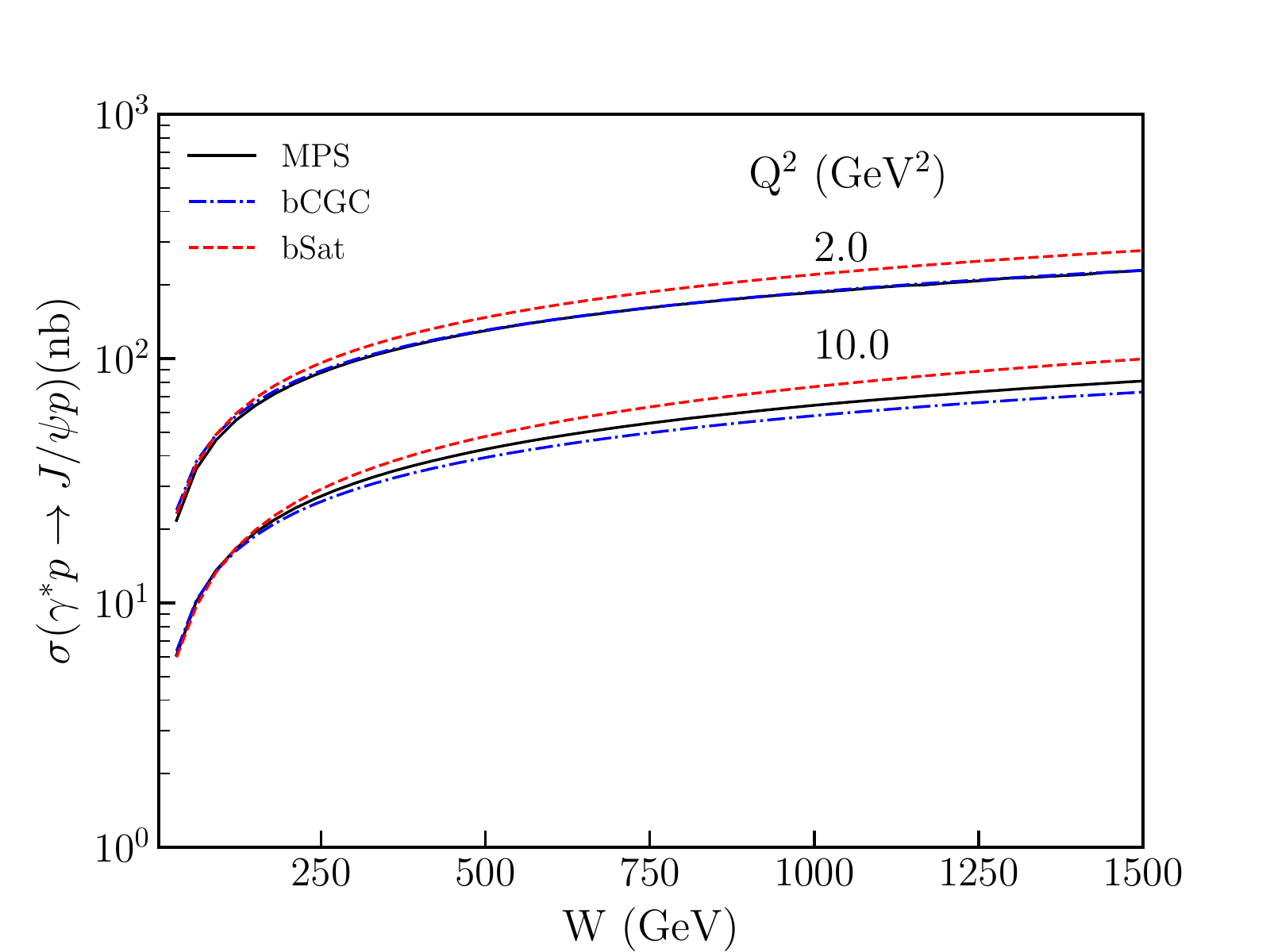}
			\includegraphics[width=0.32\textwidth]{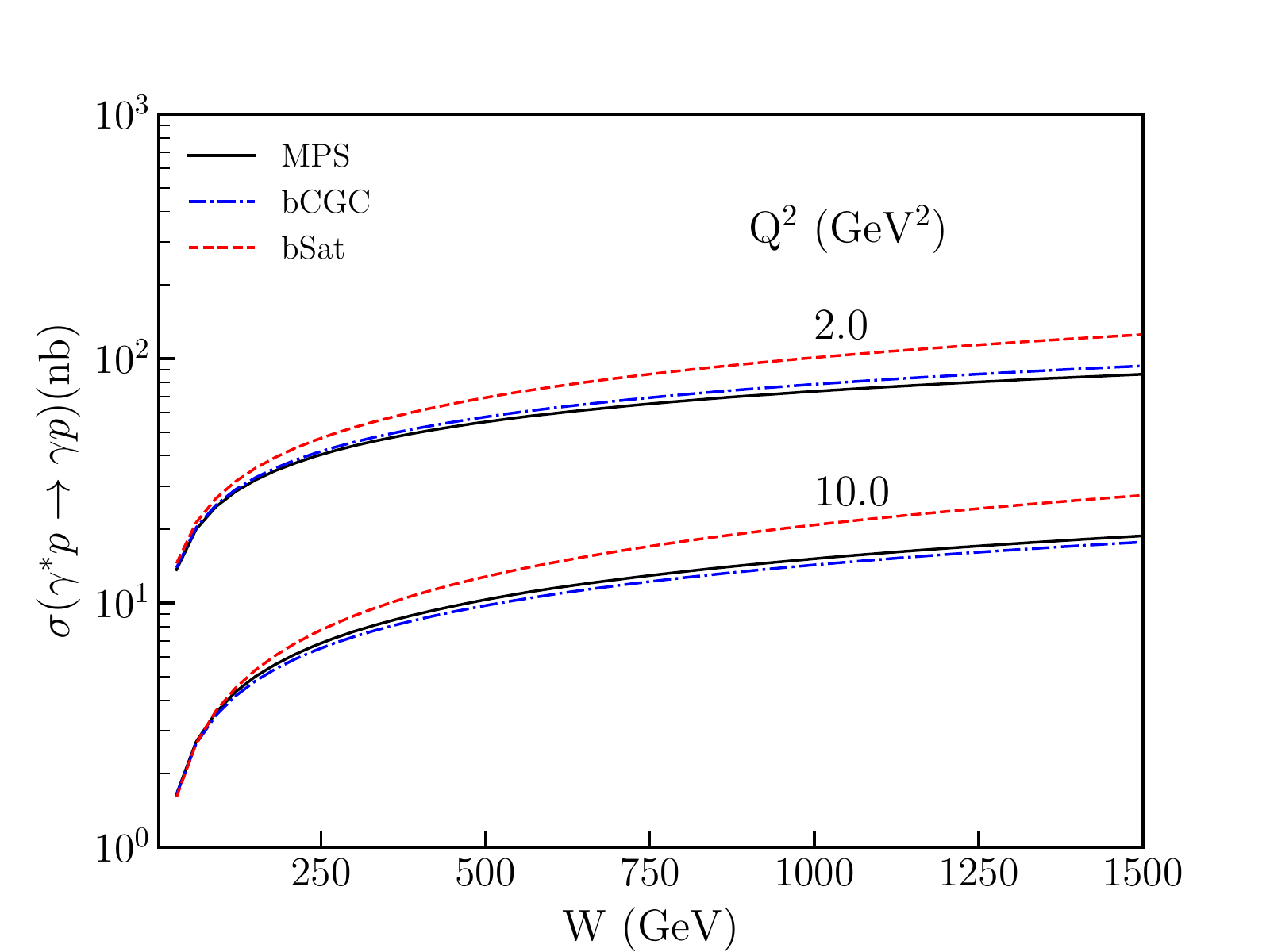}
	\caption{ Predictions of the distinct saturation models for the energy dependence of the $\rho$ (left panel), $J/\Psi$ (center panel) and DVCS (right panel) cross sections for two distinct values of $Q^2$.}
	\label{fig:energy}	
\end{figure}

As discussed in the Introduction, the search for the saturation effects is one of the major motivations for the construction of the Electron - Ion Collider (EIC) in the USA \cite{eic}, recently approved, as well as for the proposal of future electron -- hadron colliders at CERN \cite{lhec}. These colliders are expected to allow the investigation of the hadronic structure with unprecedented precision to inclusive and diffractive observables. In particular, the measurement of $d\sigma/dt$ with small error bars in a larger range of values for $t$ is expected to be feasible at the EIC and LHeC. Moreover, these future colliders are expected to  be also able to separate the events in which the virtual photon has  longitudinal or transverse polarization. Such aspects motivate the analysis of exclusive processes in the kinematical range that will be probed by the EIC and LHeC. In what follows we will perform a detailed comparison between the MPS, bSat and bCGC predictions,  considering two values for the photon virtuality ($Q^2$ = 2 and 10 GeV$^2$) and for $\gamma p$ center - of - mass energy ($W$ = 100 and 1000 GeV). Moreover, we will present, for the first time, the predictions for the $t$-distributions considering the distinct photon polarizations.
The predictions of the distinct saturation models for the energy dependence of the total cross sections are shown in Fig.~\ref{fig:energy}. One has that the MPS predictions for $\rho$ production are similar to those from the bSat one for large $W$ and small $Q^2$. For $Q^2$ = 10 GeV$^2$, the predictions of the MPS model becomes larger than those derived adopting the impact parameter saturation models. In contrast, for the $J/\Psi$ and DVCS cases, one has that the MPS and bCGC predictions are almost identical and are smaller than the bSat results. Our results indicate that in order to constrain the description of the QCD dynamics, less inclusive observables must be investigated.

\begin{figure}[t]
	\centering
	\includegraphics[width=0.48\textwidth]{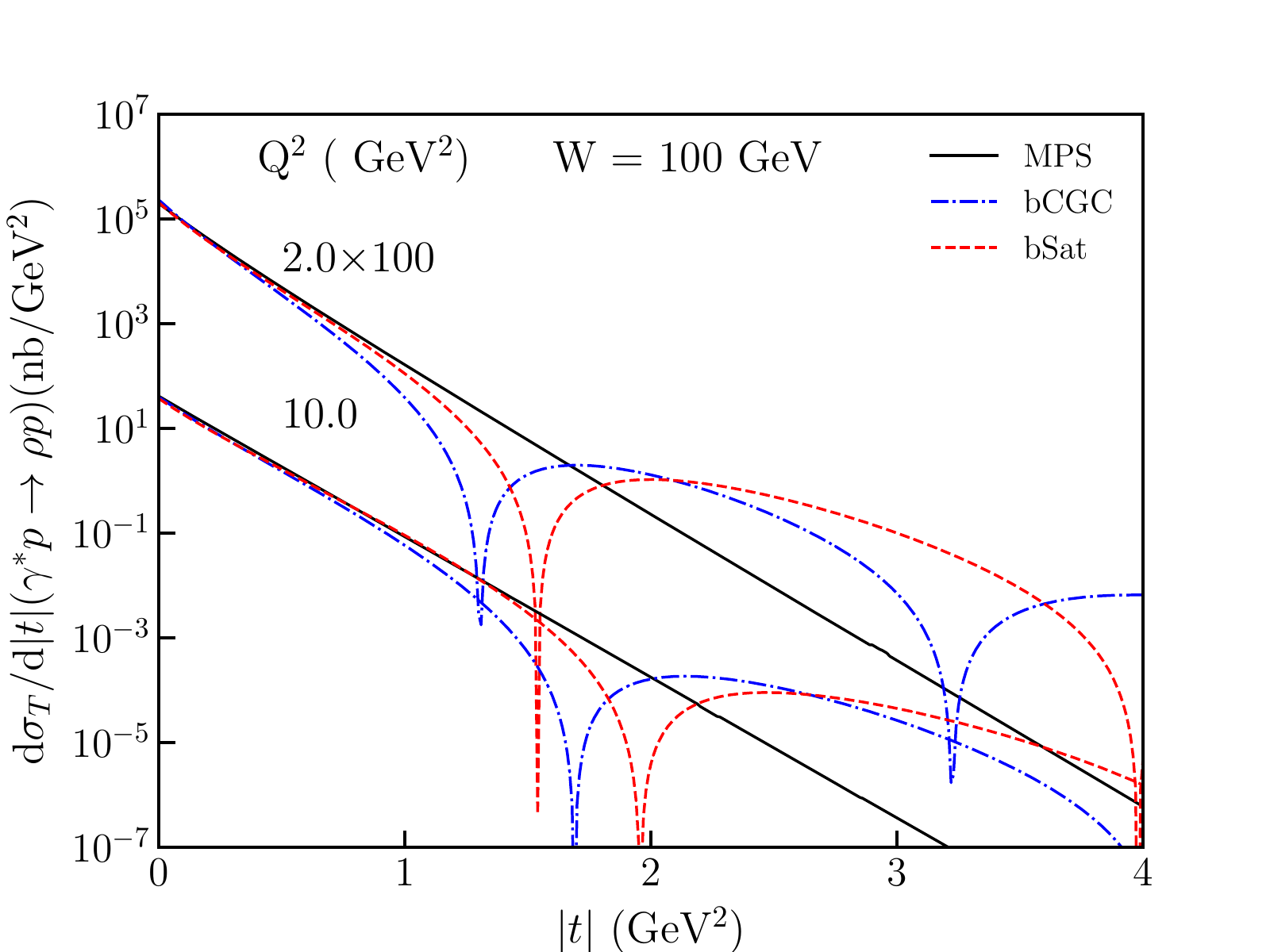}
	\includegraphics[width=0.48\textwidth]{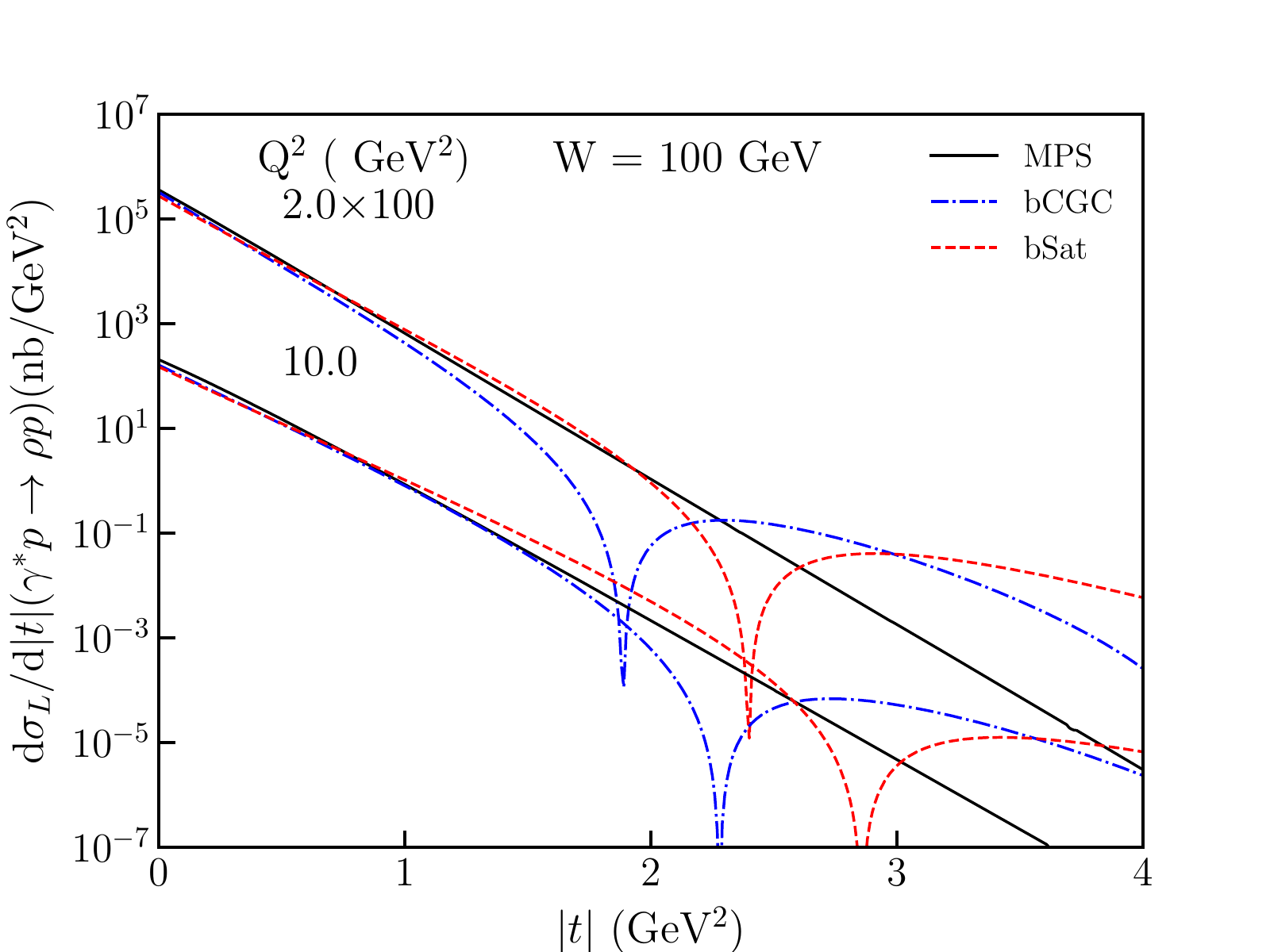} \\
		\includegraphics[width=0.48\textwidth]{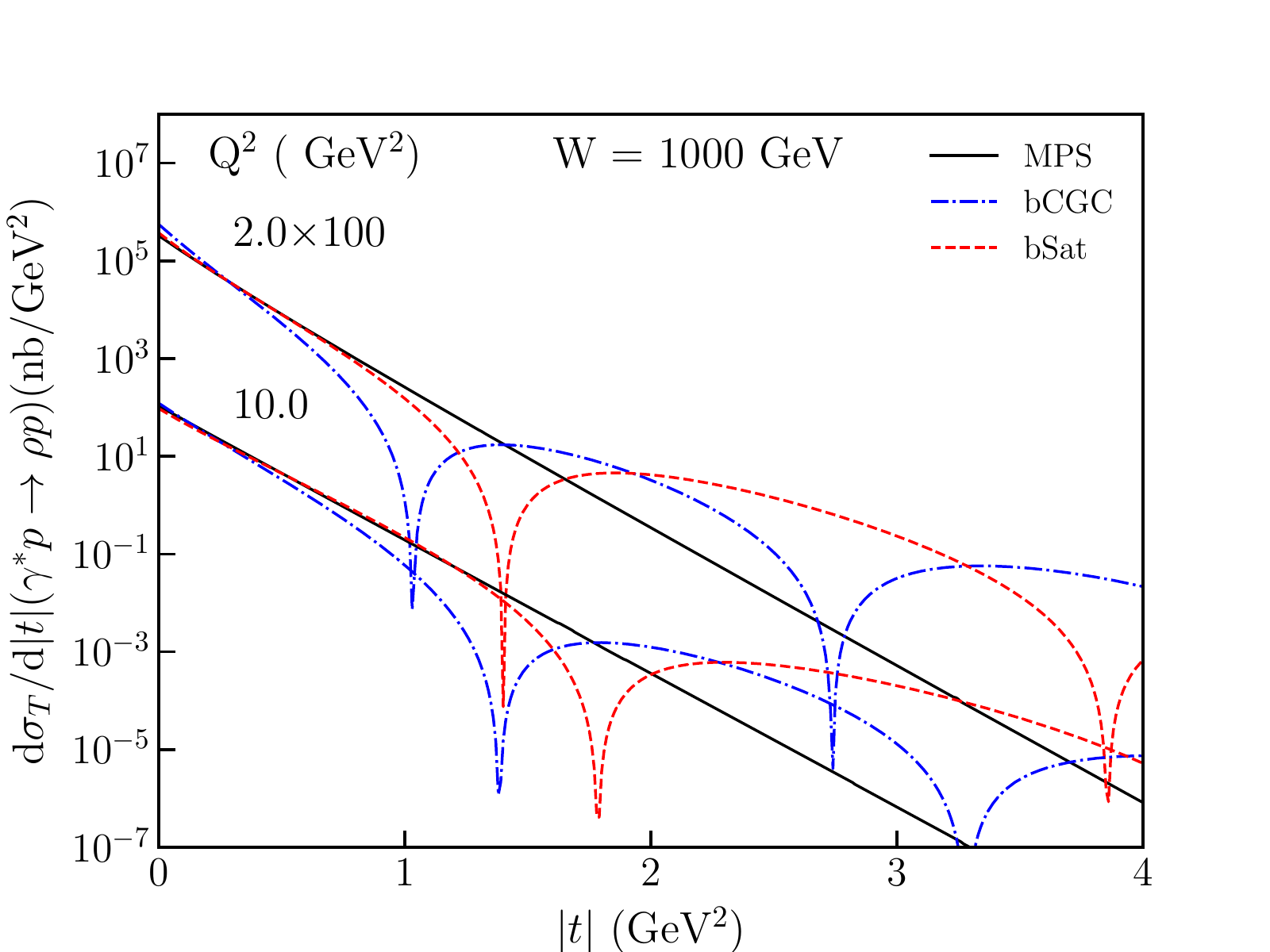}
	\includegraphics[width=0.48\textwidth]{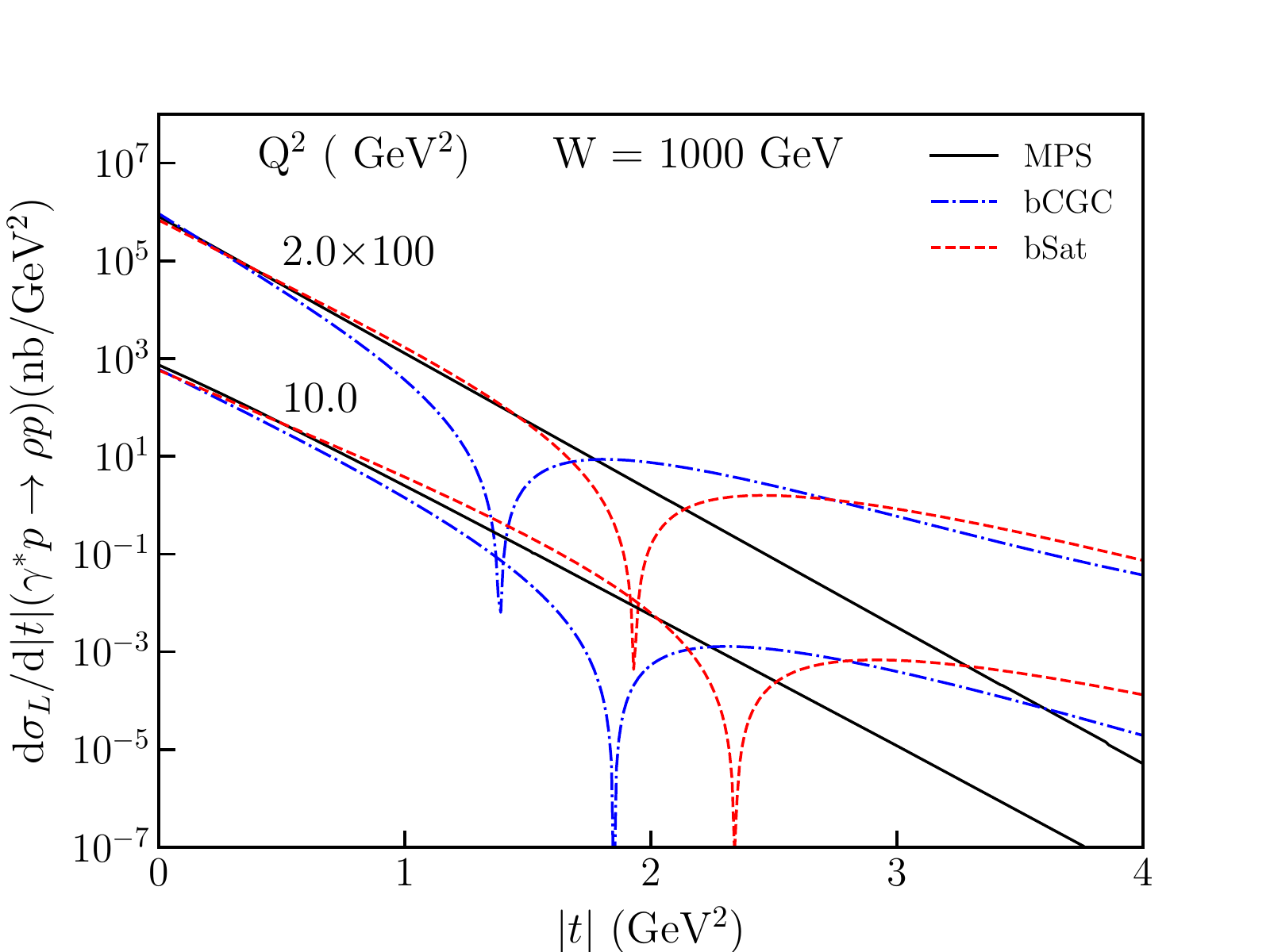}
	\caption{Predictions of the different saturation models for $d\sigma/dt$ considering the exclusive $\rho$ production for distinct values of $Q^2$ and $W$ and   that  the virtual photon has a transverse (left panels) or longitudinal (right panels) polarization. Results for $W = 100$ GeV (1000 GeV) are presented in the upper (lower) panels.  }
	\label{fig:pol_rho}	
\end{figure}

\begin{figure}[t]
		\centering
		\includegraphics[width=0.48\textwidth]{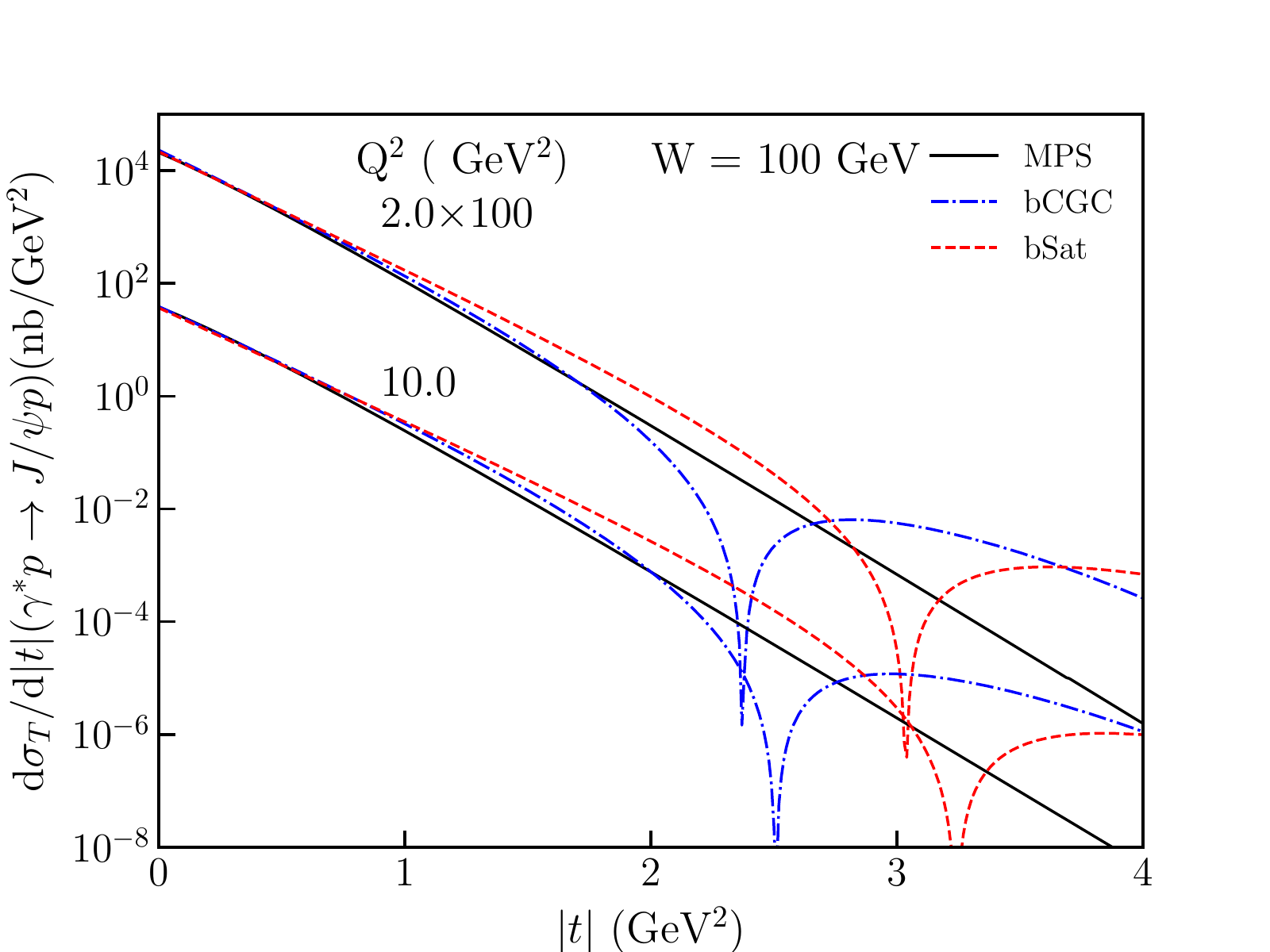}
		\includegraphics[width=0.48\textwidth]{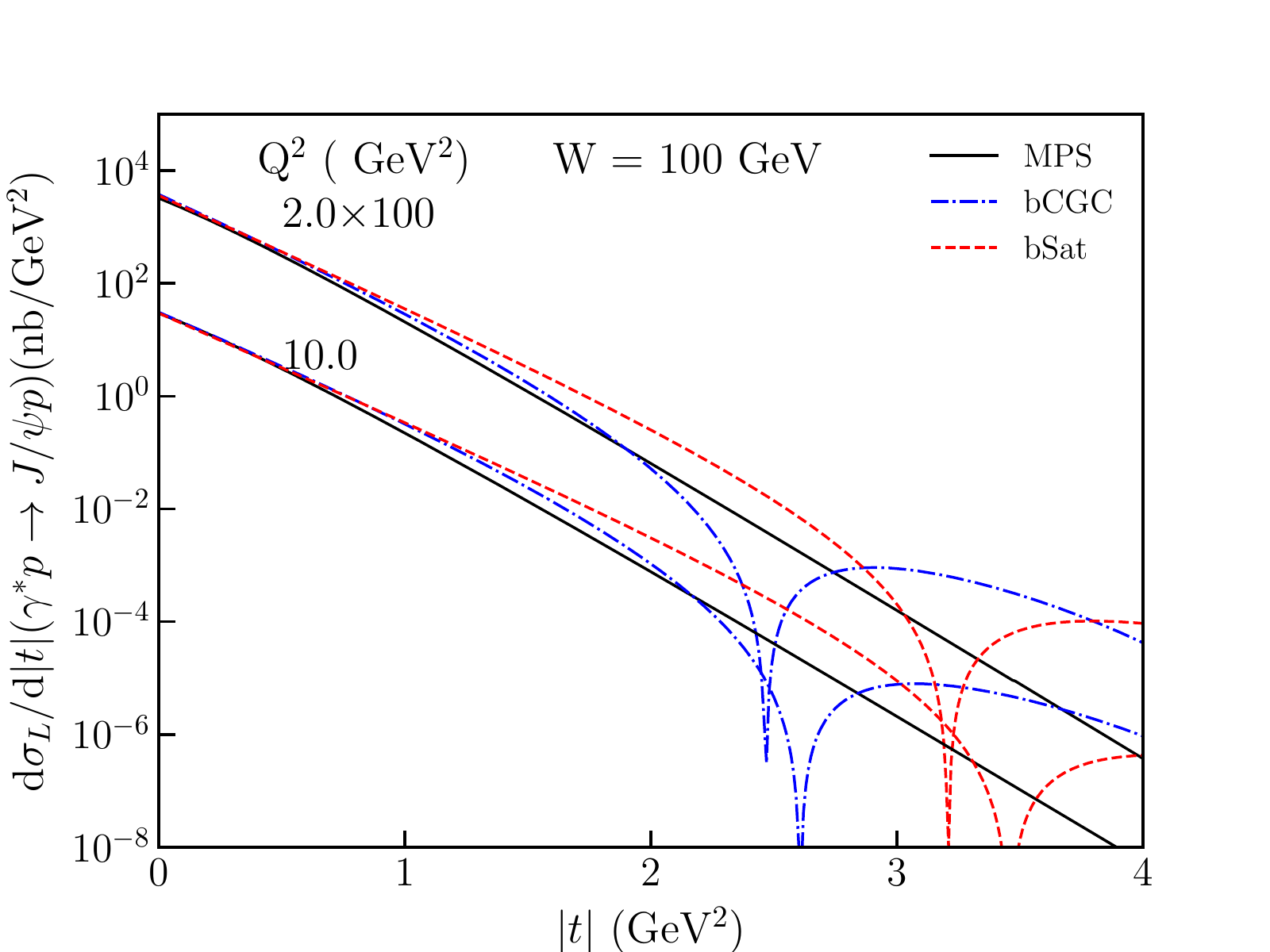} \\
			\includegraphics[width=0.48\textwidth]{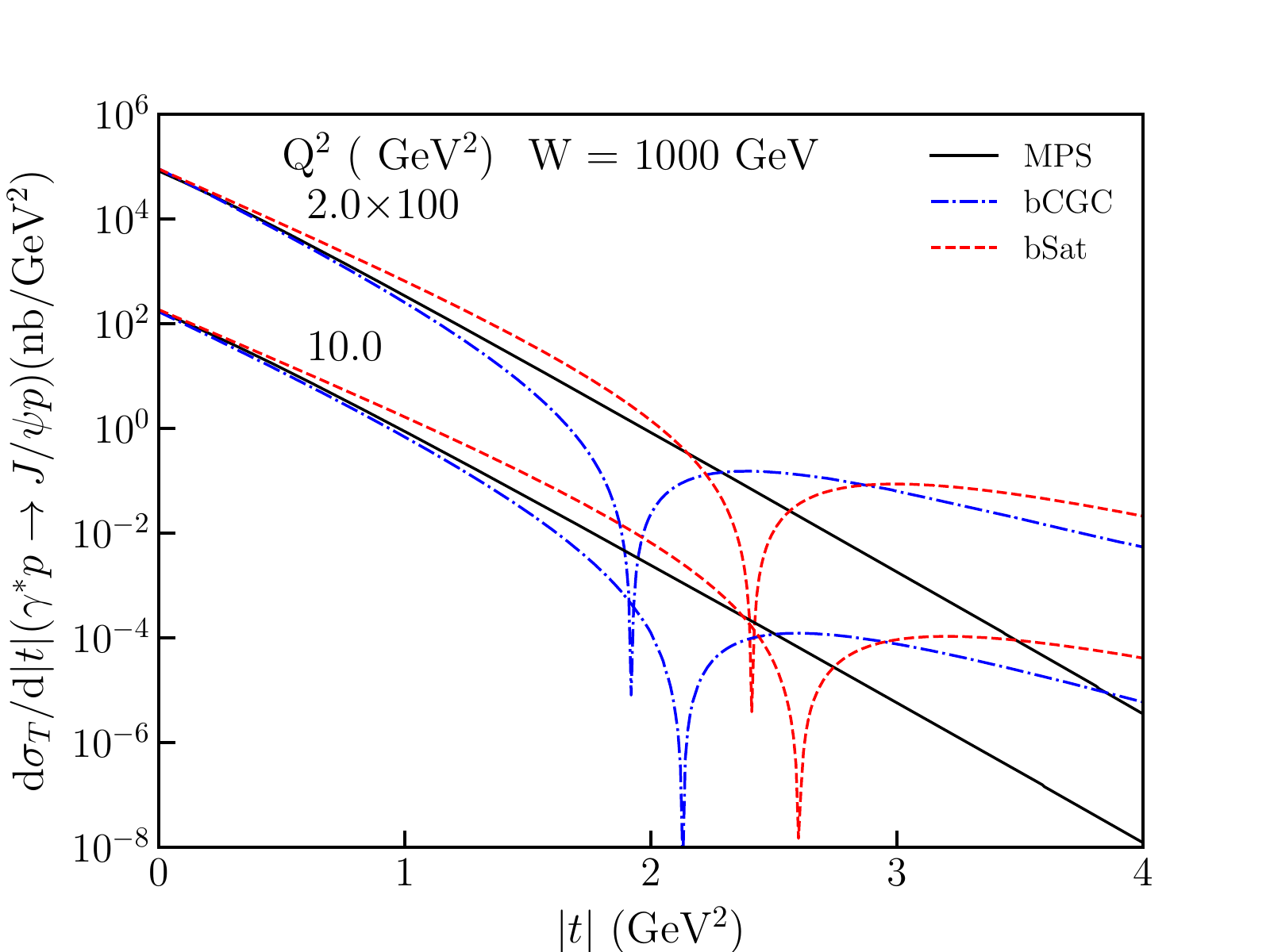}
	\includegraphics[width=0.48\textwidth]{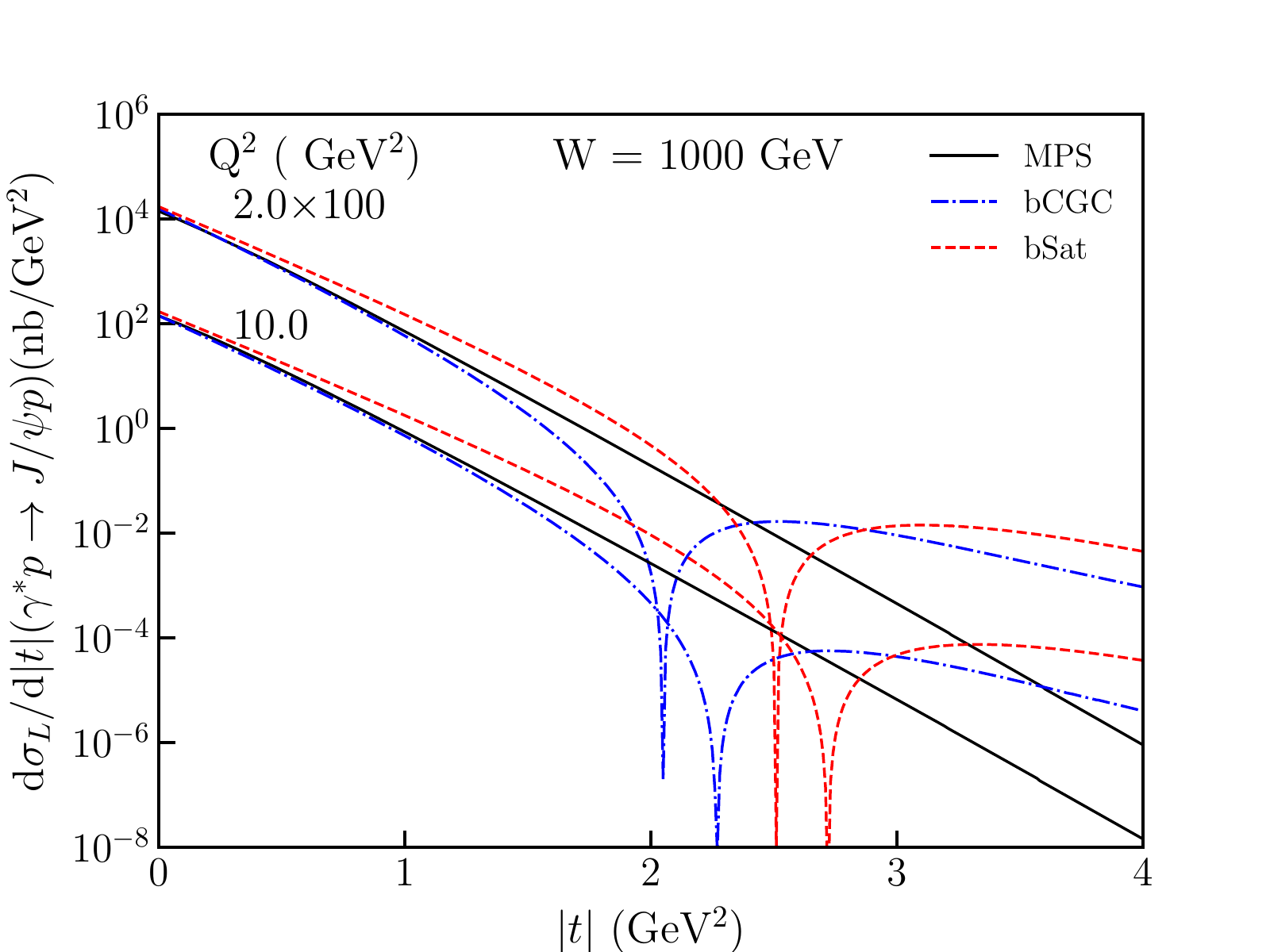}
		\caption{Predictions of the different saturation models for $d\sigma/dt$ considering the exclusive $J/\Psi$ production for distinct values of $Q^2$ and $W$ and   that  the virtual photon has a transverse (left panels) or longitudinal (right panels) polarization. Results for $W = 100$ GeV (1000 GeV) are presented in the upper (lower) panels.  }
		\label{fig:pol_jpsi}		
\end{figure}

In Figs.~\ref{fig:pol_rho} and \ref{fig:pol_jpsi} we display our predictions for the exclusive $\rho$ and $J/\Psi$ production, respectively, considering that the virtual photon has a transverse (left panels) or longitudinal (right panels) polarization. Results for $W$ = 100 GeV (1000 GeV) are presented in the upper (lower) panels.  
One has that the distinct predictions are similar for $|t| \approx 0$, which is expected from the results obtained in Fig.~\ref{fig:energy}. However, they significantly differ at large values of the squared transferred momentum.
One has that while the bSat and bCGC models predict the presence of dips in the distribution, these are not predicted by the MPS one in the kinematical range considered, independent of the photon polarization. Such result is directly associated to the factorization of  the form factor $f(\qb)$ in the MPS model. In contrast, for the bSat and bCGC models, one also has  that the position of the dip is dependent of the dipole model, with the first dip predicted by the bCGC model occuring for smaller values  of $|t|$ in comparison to the bSat prediction. Moreover,   the first dip also occurs for smaller values of $|t|$ when the photon virtuality is decreased or the center - of - mass energy is increased. Another important aspect is that the position of the dips is  dependent of the photon polarization, with the dip predicted by the bSat and bCGC models occuring for larger values of $|t|$ in the case of a photon with a longitudinal polarization. In particular, for the $\rho$ production, we predict a large difference between the positions of the dips for longitudinal and transverse polarizations.  Such result indicates that a future experimental analysis of $d\sigma/dt$ for the distinct photon polarizations will be useful to discriminate between the distinct phenomenological approaches for saturation physics. 

%%%%%%%%%%%%%%%%%%%%%%%%%%%%%%%%%%%%%%%%%%%%%%%%%%%%%%%%%%%%%%%%%%%%%%%%%%%%%%%%%

\begin{figure}[t]
	\centering
	\includegraphics[width=0.48\textwidth]{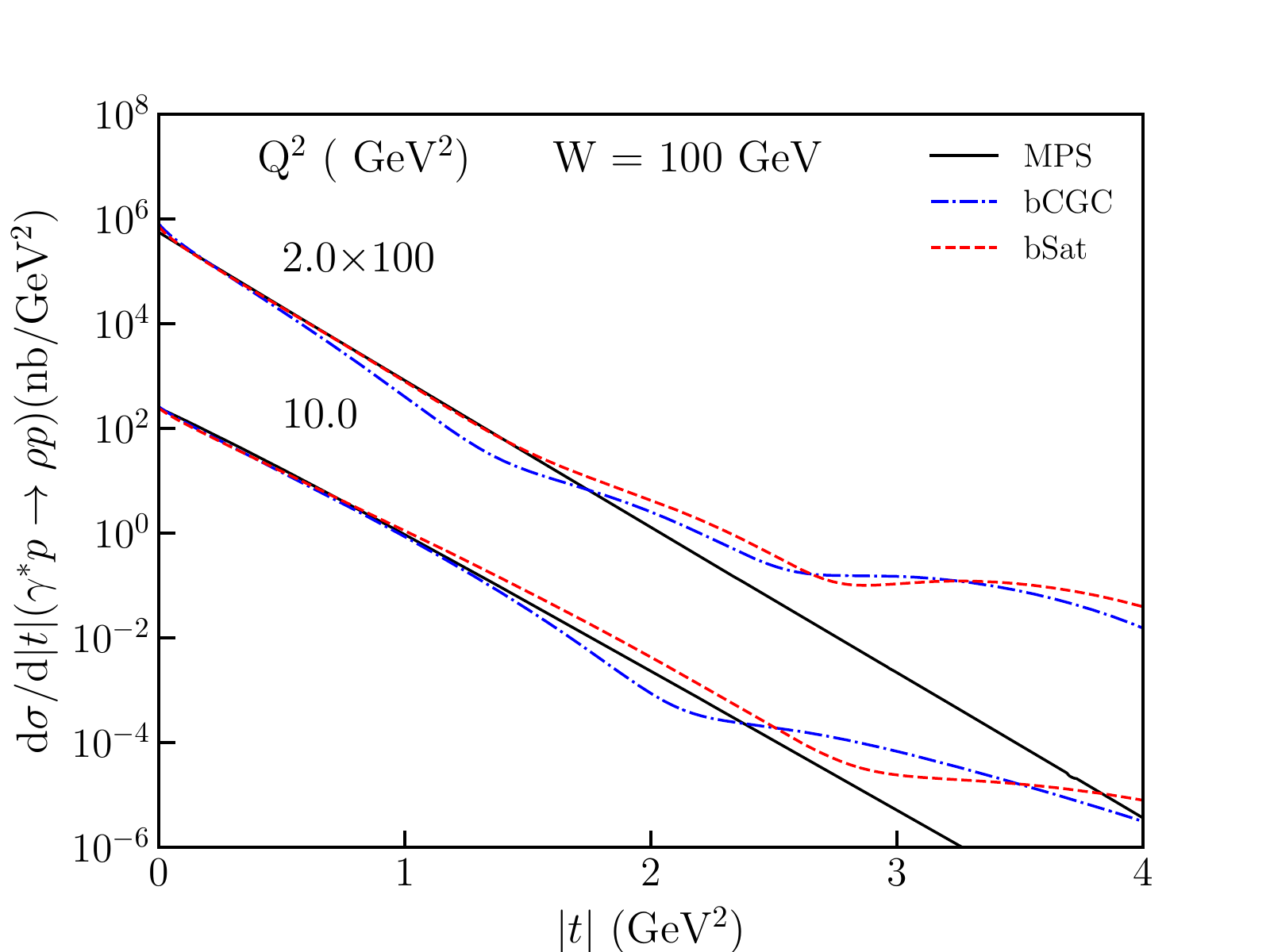}
	\includegraphics[width=0.48\textwidth]{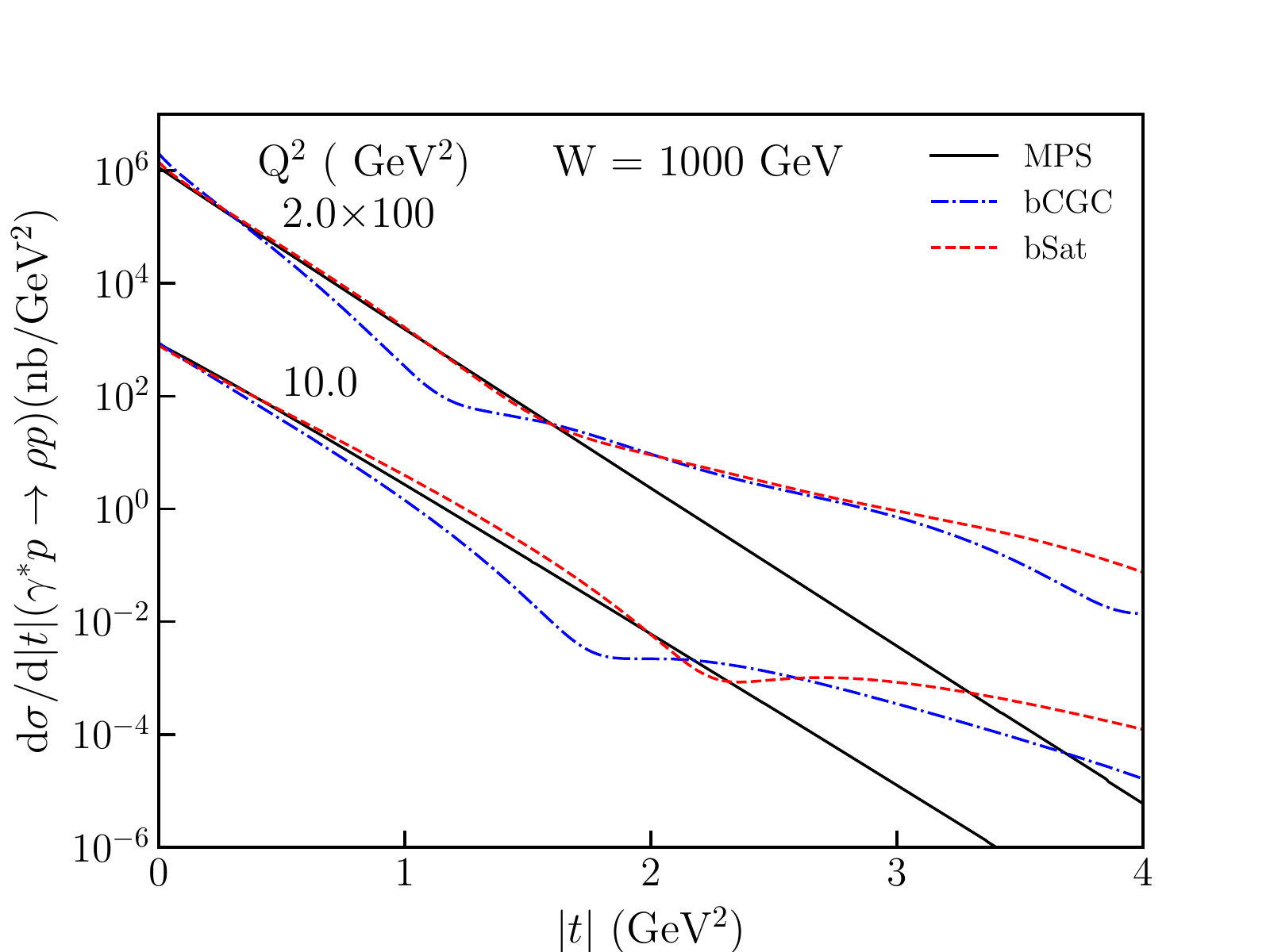} \\
	\includegraphics[width=0.48\textwidth]{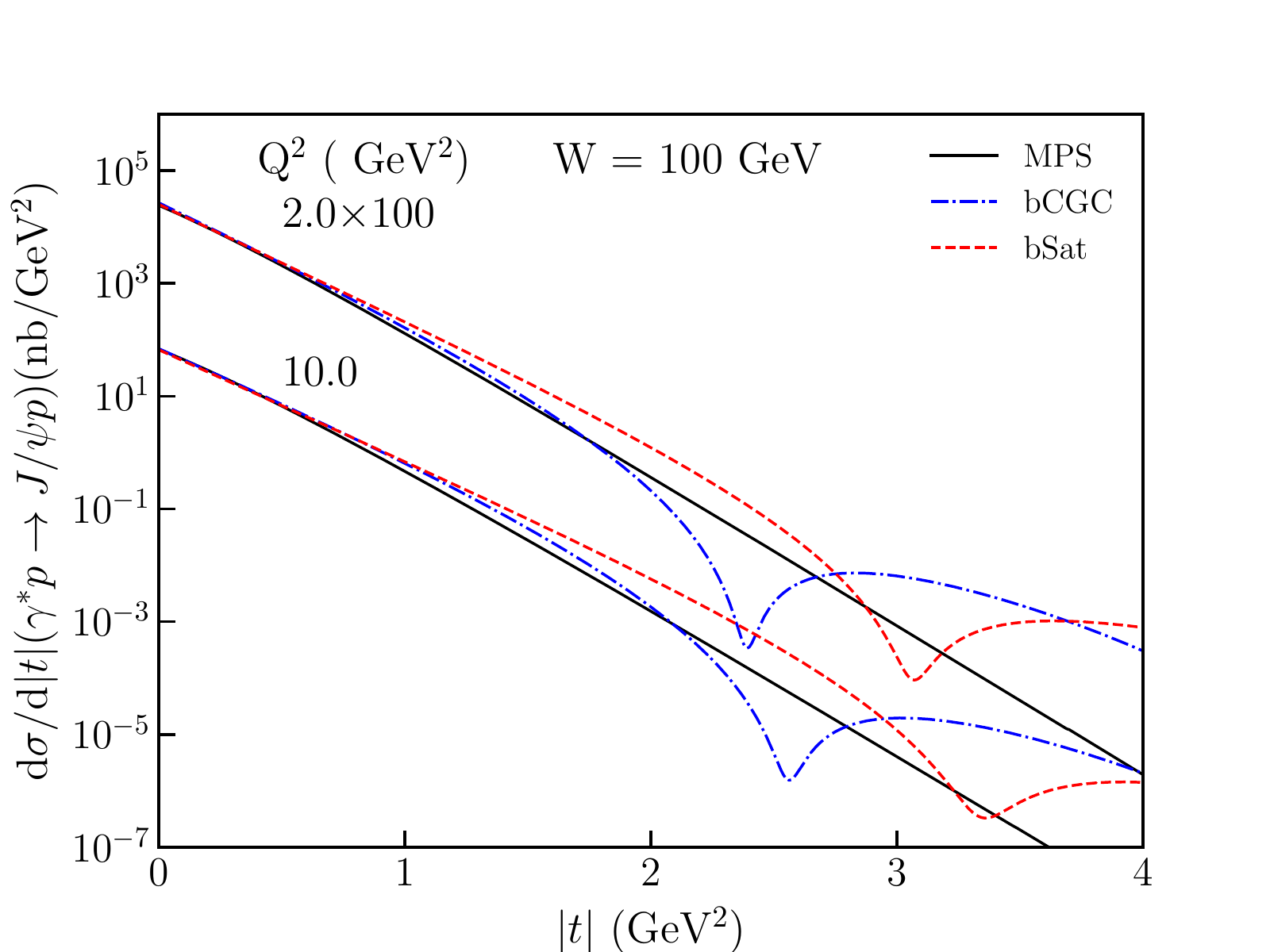}
	\includegraphics[width=0.48\textwidth]{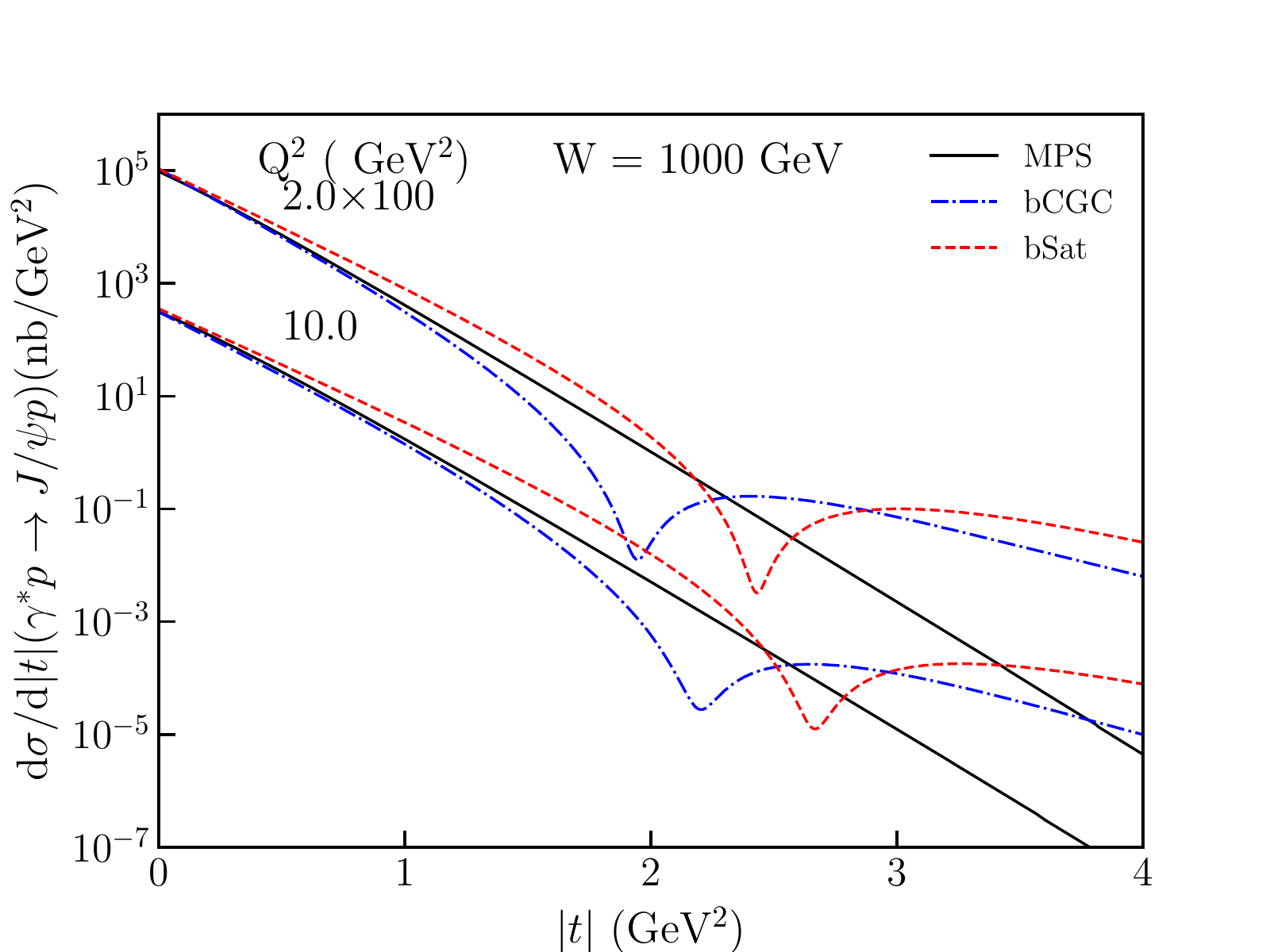} \\
	\includegraphics[width=0.48\textwidth]{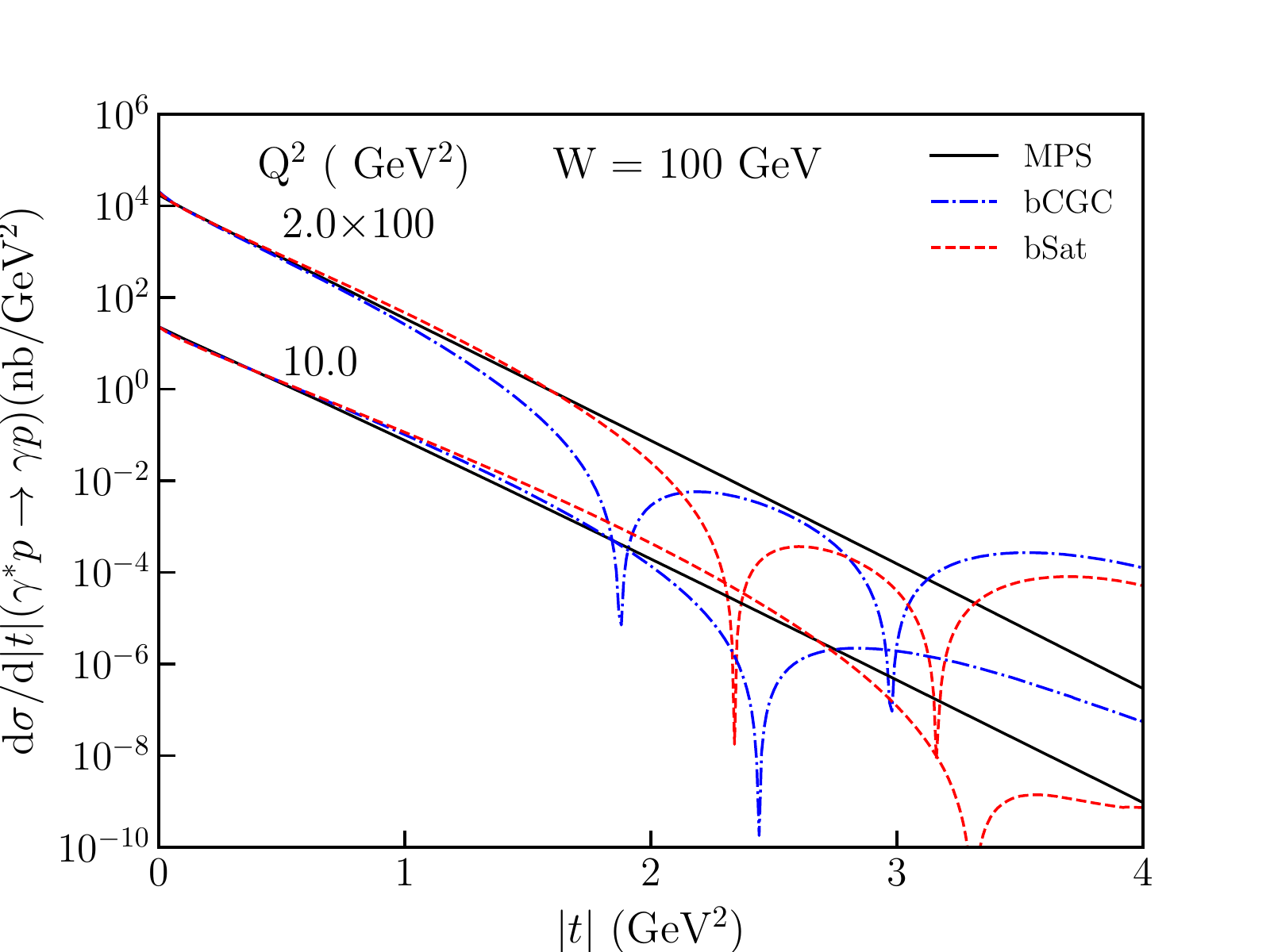}
	\includegraphics[width=0.48\textwidth]{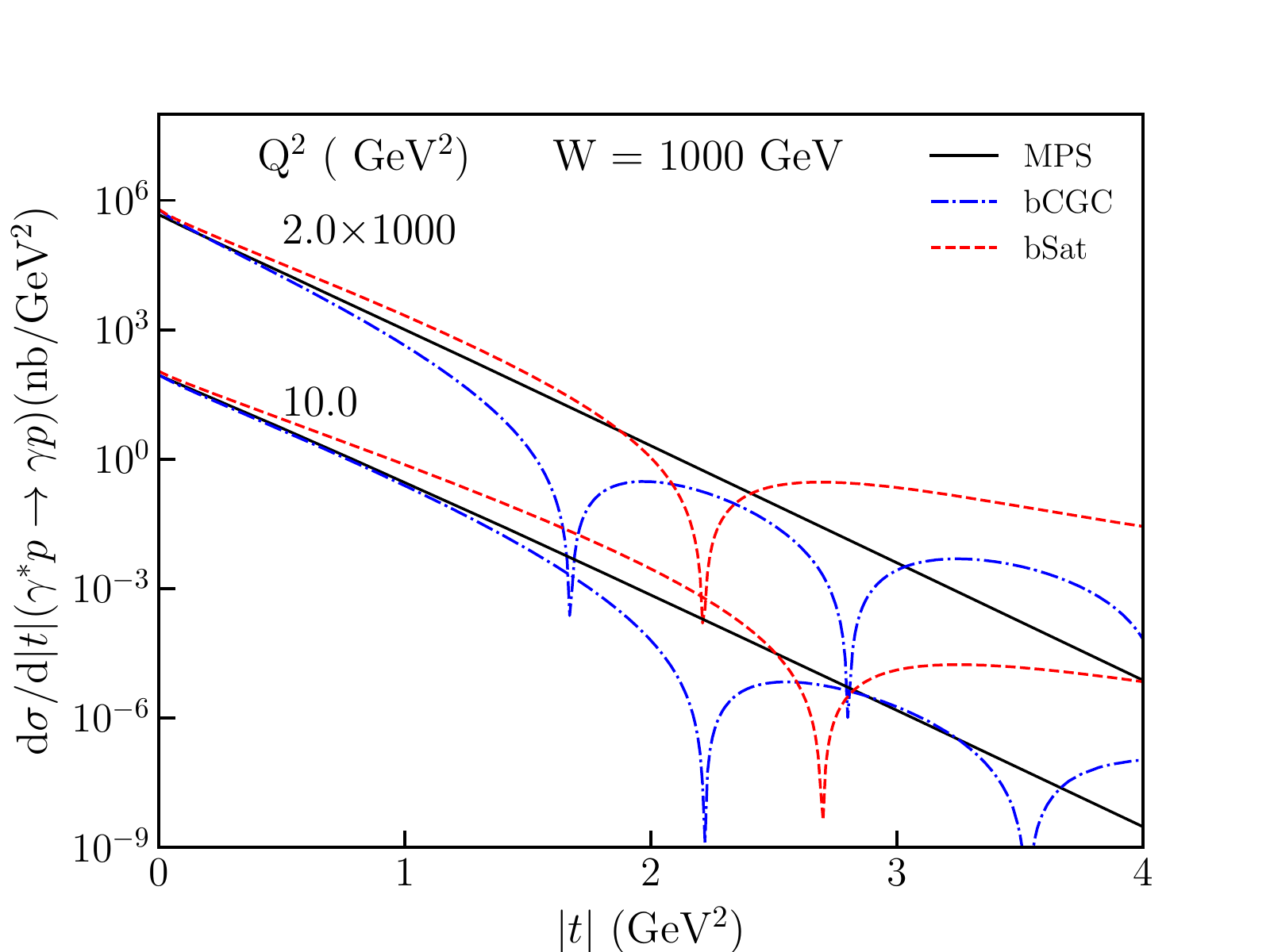}
	\caption{Predictions of the different saturation models for $d\sigma/dt$, summed over photon polarizations, considering the exclusive $\rho$ (upper),  $J/\Psi$ (middle) and $\gamma$ (lower) production for distinct values of $Q^2$ and $W$. Results for $W = 100$ GeV (1000 GeV) are presented in the left (right) panels. }
	\label{fig:total}	
\end{figure}

In Fig.~\ref{fig:total} we exhibit our predictions   
for $d\sigma/dt$ considering the sum over photon polarizations. Results for the exclusive $\rho$ (upper),  $J/\Psi$ (middle) and $\gamma$ (lower) production are presented considering distinct values of $Q^2$ and $W$.
One has that when the contribution of both photon polarizations are added, the shape of the differential distributions for $\rho$ and $J/\Psi$ predicted by the bSat and bCGC models are modified. In particular, for the $\rho$ production, where the difference between the position of the dips in the longitudinal and transverse polarizations is not negligible, one has that the diffraction pattern predicted by these models almost vanishes. As a consequence, the predictions of the distinct saturation models become similar. For the $J/\Psi$ case, one has the diffraction pattern is still present, which implies that the analysis of the total  $d\sigma/dt$ is also a good observable to discriminate between the distinct approaches. Finally, for the DVCS process (lower panels), which is characterized by incoming photons of transverse polarization, one has that the predictions of the different saturation models are very distinct, which indicate that a future experimental analysis of the exclusive photon production,  considering events characterized by squared transferred momentum $|t|$ larger than 1.8 GeV$^2$, will be able to constrain the description of the QCD dynamics.

\section{Summary}
\label{sec:conc}
During the last years, several studies have shown that the exclusive production of vector mesons and photons  in $ep$ collisions have the  
potential to probe the QCD dynamics at high energies. In particular, there is the expectation that the large data sample to be obtained at the EIC and LHeC  will allow a more detailed study of the exclusive cross sections  and  a better discrimination between 
alternative descriptions. 
In this paper we have presented a comprehensive study of the exclusive $\rho$, $J/\Psi$ and $\gamma$ production  in $ep$ collisions at future colliders  
using the  color dipole formalism.  We have used three  different 
models for the dipole scattering amplitude, which take into account   the 
non - linear (saturation) effects of the QCD dynamics. In particular, we have updated the MPS model, which is based on the solutions of the BK equation in the momentum space, and is characterized by a saturation scale that is dependent of the squared momentum transfer $t$ and by the factorization of the $t$ - dependence associated to the proton vertex. A very good description of the current data is obtained when the Light Cone Gauss model is assumed for the vector meson wave functions. A detailed comparison of the MPS predictions with the bSat and bCGC results is performed. Differently from the bSat and bCGC model, a diffractive pattern  in $d\sigma/dt$ is not predicted by the MPS model. Moreover, one has obtained that the bSat and bCGC predictions for the position of the dips are distinct and dependent on the virtual photon polarization. Our results indicate that a future experimental analysis of  
exclusive processes,  considering events characterized by large values of the squared transferred momentum, has the potentiality of  constraining the description of the QCD dynamics.

\section*{Acknowledgments}
The work is partially supported by the Strategic Priority Research Program of Chinese Academy of Sciences (Grant NO. XDB34030301). 
VPG was  partially financed by the Brazilian funding
agencies CNPq,   FAPERGS and  INCT-FNA (process number 
464898/2014-5).

\end{document}